 \documentclass[final,3p,times]{elsarticle}
\usepackage{amssymb}
\usepackage{amssymb,amsmath,amscd,mathrsfs,amsbsy,amsfonts}
\usepackage{pstricks,pst-node}
\usepackage[metapost]{mfpic}
\usepackage{mflogo}
\usepackage{graphicx}
\usepackage{epstopdf}
\usepackage{amscd}
\usepackage{graphics}

\newtheorem{Theorem}{Theorem}[section]

\numberwithin{equation}{section}

\numberwithin{figure}{section}

\def\sech{\mathop{\rm sech}\nolimits}
\renewcommand\labelitemi{\ifmmode\circ\else$\circ$\fi}
\newcommand{\lyxdot}{.}

\journal{Mathematics and Computers in Simulation}

\begin{document}
%\begin{frontmatter}

\title{Numerical study of the KP equation for non-periodic waves}

\author{Chiu-Yen Kao$^\dagger$ and Yuji Kodama$^*$}
\address{Department of Mathematics, The Ohio State University, Columbus,
OH 43210}
%\email{kodama@math.ohio-state.edu}

\thanks{$^\dagger$ Partially supported by NSF grant DMS,  $^*$ Partially
supported by NSF grant DMS0806219}

%\maketitle

\begin{abstract}
The Kadomtsev-Petviashvili (KP) equation describes weakly dispersive
and small amplitude waves propagating in a quasi-two dimensional
situation. Recently a large variety of exact soliton solutions of
the KP equation has been found and classified. Those
soliton solutions are localized along certain lines in a
two-dimensional plane and decay exponentially everywhere else, and
they are called line-soliton solutions in this paper. The
classification is based on the far-field patterns of the solutions
which consist of a finite number of line-solitons. In this paper, we
study the initial value problem of the KP equation with V- and
X-shape initial waves consisting of two distinct line-solitons by
means of the direct numerical simulation. We then show that the
solution converges asymptotically to some of those exact soliton
solutions. The convergence is in a locally defined
$L^2$-sense. The initial wave patterns
considered in this paper are related to the rogue waves generated by
nonlinear wave interactions in shallow water wave problem.
 \end{abstract}


%\end{frontmatter}

\maketitle
%\tableofcontents
%\thispagestyle{empty}

%%%%%%%%%%%%%%%%%%%%%%%%%%%%%%%%%%%%%%%%%%%%%%%%%%%%%%%%%%%%%%%%%%%%%%%%%%%%%%%
\section{Introduction}
The KdV equation may be obtained in the leading order approximation
of an asymptotic perturbation theory for one-dimensional nonlinear
waves under the assumptions of weak nonlinearity (small amplitude)
and weak dispersion (long waves). The initial value problem of the
KdV equation has been extensively studied by means of the method of
inverse scattering transform (IST). It is then well-known that a
general initial data decaying rapidly for large spatial variable
evolves into a sum of individual solitons and some weak dispersive
wave trains separated away from solitons (see for examples,
\cite{AS:81, N:85, NMPZ:84, Wh:74}).

In 1970, Kadomtsev and Petviashvili \cite{KP:70} proposed a two-dimensional dispersive wave equation
to study the stability of one soliton solution of the KdV equation under the influence of
weak transversal perturbations. This equation is now referred to as the KP equation, and
considered to be a prototype of the integrable nonlinear dispersive wave equations in {{two dimensions}}.
The KP equation can be also represented in the Lax form, that is, there exists a pair of linear
equations associated with the eigenvalue problem and the evolution of the eigenfunctions.
However, unlike the case of the KdV equation, the method of IST based on the pair of linear equations
does not seem to provide a practical method for the initial value problem with {{non-periodic}} waves considered in this paper.  At the present time, there is no feasible analytic method to solve the initial value problem of the KP equation with initial waves having line-solitons in the far field.

In this paper, we study this type of the initial value problem of the KP equation by means of the
direct numerical simulation.  In particular, we consider the following two cases of the initial
waves:  In the first case, the initial wave consists of two semi-infinite line-solitons forming a V-shape
pattern, and in the second case, the initial wave is given by a linear combination of
two infinite line-solitons forming X-shape.  Those initial waves have been considered in
the study of the generation of large amplitude waves in shallow water
\cite{PTLO:05,TO:07,KOT:09}.  The main result of this paper is to show that the solutions of the initial value problem with {{those initial waves}} asymptotically {\it converge} to
some of the exact soliton solutions found in \cite{CK:08, CK:09}.
This implies a separation of the (exact) soliton solution from the dispersive radiations in the similar manner
as in the KdV case.

The paper is organized as follows:  In Section \ref{S:KP}, we provide a brief summary of the soliton solutions of the KP equation and the classification theorem obtained in \cite{CK:08, CK:09} for those soliton solutions as a background necessary for the present study. In particular, we introduce the parametrization of each soliton solution with a {\it chord diagram} which represents
a derangement of the permutation group, i.e. permutation without fixed point.
In Section \ref{sec:exact}, we present several exact soliton solutions, and describe some
properties of those solutions. Each of those soliton solutions has $N_+$ numbers of  line-solitons
in a far field on the two-dimensional plane, say in $y\gg 0$,  and $N_-$ numbers of line-solitons in the far field of the opposite side, i.e. $y\ll 0$. This type of soliton solution is referred to as an $(N_-,N_+)$-soliton solution.
Here we consider those solitons with $N_-+N_+\le 4$ and $N_-\le 3$.
In Section \ref{sec:numerics}, we describe the numerical scheme used in this paper,
which is based on the pseudo-spectral method combined with the window technique \cite{S:05, TOM:08}.  The window technique is especially used  to compute our non-periodic problem which is
essentially an infinite domain problem.
Finally, in Section \ref{sec:result}, we present the numerical results of the initial value problems with
V- and X-{{shape}} initial waves, and show that the solutions asymptotically converge to some of
those exact solutions discussed in Section \ref{sec:exact}.  The convergence is in the sense of locally defined
$L_2$-sense with the usual norm, i.e.
$
\| f\|_{L^2(D)}:=\left(\iint_D|f(x,y)|^2 dxdy\right)^{\frac{1}{2}},
$
where $D\subset\mathbb{R}^2$ is a compact set which covers the main structure describing the
(resonant) interactions in the solution.
We also propose a method to identify an exact solution for a given initial wave with
 V- or X-shape pattern based on the chord diagrams introduced in the classification theory.

\section{Background}\label{S:KP}
Here we give a brief summary of the recent result of the classification theorem
for soliton solutions of the KP equation (see \cite{K:04,CK:08,CK:09} for the details). In particular, each soliton solution
is then parametrized by a chord diagram which represents a unique element of the permutation group. We use this parametrization
throughout the paper.

\subsection{The KP equation}

The KP equation is a two-dimensional nonlinear dispersive wave equation given by
\begin{equation}\label{kp}
{\partial_x}\left(4{\partial_t}u+6u{\partial_x}u+{\partial_x^3}u\right)+3{\partial_y^2}u=0,
\end{equation}
where $\partial_x^nu:=\displaystyle{\frac{\partial^nu}{\partial x^n}}$ etc.
Let us express the solution in the form,
\begin{equation}\label{tau}
u(x,y,t)=2\,{\partial_x^2}\,\ln\tau(x,y,t)\,.
\end{equation}
where the function $\tau$ is called the {\it tau} function, which plays
a central role in the KP theory.
In this paper,
we consider a class of the solutions, where each solution can be expressed by $\tau(x,y,t)$ in the Wronskian
determinant form $\tau={\rm Wr}(f_1,\ldots,f_N)$, i.e.
\begin{equation}\label{wronskian}
\tau(x,y,t)=\left|\begin{matrix}
f_1 &  f_1' &\cdots & f_1^{(N-1)} \\
f_2 & f_2' & \cdots & f_2^{(N-1)}\\
\vdots &\vdots & \ddots & \vdots \\
f_N & f_N' &\cdots &f_N^{(N-1)}
\end{matrix}\right|\,,
\end{equation}
with $f_n^{(j)}:={\partial_x^j}f_n$ for $n=1,\ldots,N$. Here the
functions $\{f_1,\ldots,f_N\}$ form a set of linearly independent
solutions of the linear equations,
\[
{\partial_y}f_n={\partial_x^2}f_n,\qquad {\partial_t}f_n=-{\partial_x^3}f_n.
\]
(The fact that (\ref{tau}) with (\ref{wronskian}) gives a solution of the KP equation is well-known and
the proof can be found in several places, e.g. see \cite{H:04, CK:09}.)
The solution of those equations can be expressed  in the Fourier transform,
\begin{equation}\label{Fourier}
f_n(x,y,t)=\int_Ce^{kx+k^2y-k^3t}\,\rho_n(k)\,dk\,, \qquad n=1,2,\ldots,N,
\end{equation}
with an appropriate contour $C$ in $\mathbb{C}$ and the measure $\rho_n(k)\,dk$.
In particular, we
consider a finite dimensional solution with $\rho_n(k)\,dk=\sum_{m=1}^M\,a_{n,m}\delta(k-k_m)\,dk$ with $a_{n,m}\in\mathbb{R}$, i.e.
\[
f_n(x,y,t)=\sum_{m=1}^Ma_{n,m}E_m(x,y,t)\qquad {\rm with}\quad E_m=\exp(k_mx+k_m^2y-k_m^3t).
\]
Thus this type of solution is characterized by the parameters $\{k_1,k_2,\ldots,k_M\}$ and
the $N\times M$ matrix $A:=(a_{n,m})$ of rank$(A)=N$, that is, we have
\begin{equation}\label{NM}
(f_1,f_2,\ldots,f_N)=(E_1,E_2,\ldots, E_M)A^T\,.
\end{equation}
Note that $\{E_1,E_2,\ldots,E_M\}$ gives a basis of $\mathbb{R}^M$ and $\{f_1,f_2,\ldots,f_N\}$
spans an $N$-dimensional subspace of $\mathbb{R}^M$. This means that the $A$-matrix
can be identified as a point on the real Grassmann manifold Gr$(N,M)$ (see \cite{K:04,CK:09}).
More precisely, let $M_{N\times M}({\mathbb R})$ be the set of all $N\times M$ matrices of rank $N$.
Then Gr$(N,M)$ can be expressed as
\[
{\rm Gr}(N,M)={\rm GL}_N({\mathbb R})\backslash M_{N\times M}({\mathbb R}),
\]
where GL$_N({\mathbb R})$ is the general linear group of rank $N$. This is saying that
other basis $(g_1,\ldots,g_N)=(f_1,\ldots,f_N)H$ for any $H\in{\rm GL}_N({\mathbb R})$
spans the same subspace. Notice here that the freedom in the $A$-matrix with GL$_N({\mathbb R})$ can be fixed by expressing $A$ in the reduced row echelon form (RREF).
We then assume throughout this paper that the $A$-matrix is in the RREF, and  show that
the $A$-matrix plays a crucial role in our discussion on the asymptotic behavior of the initial value problem.

Now using the Binet-Cauchy Lemma for the determinant, the $\tau$-function of (\ref{tau})
can be expressed in the form,
\begin{align}
\tau&=\left|\begin{pmatrix}
E_1 & E_2 &\cdots &\cdots& E_M\\
k_1E_1& k_2E_2 & \cdots & \cdots&k_ME_M\\
\vdots &\vdots & \ddots & \ddots &\vdots \\
k_1^{N-1}E_1& k_2^{N-1}E_2&\cdots&\cdots &k_M^{N-1}E_M
\end{pmatrix}\begin{pmatrix}
a_{11} & a_{21}&\cdots &a_{N1}\\
a_{12}&a_{22} & \cdots &a_{N2}\\
\vdots &\vdots &\ddots & \vdots\\
\vdots&\vdots&\ddots&\vdots\\
a_{1M}& a_{2M}&\cdots &a_{NM}
\end{pmatrix}\right| \nonumber \\
&=\sum_{1\le j_1<j_2<\cdots< j_N\le M}\xi(j_1,j_2,\ldots,j_N)E(j_1,j_2,\ldots,E_N), \label{B-C}
\end{align}
where $\xi(j_1,\ldots,j_N)$ is the $N\times N$ minor of the $A$-matrix with $N$ columns
marked by $(j_1,\ldots,j_N)$, and $E(j_1,\ldots,j_N)$ is given by
\[
E(j_1,\ldots,j_N)={\rm Wr}(E_{j_1},\ldots,E_{j_N})=\prod_{l<m}(k_{j_m}-k_{j_l})E_{j_1}\cdots E_{j_N}.
\]
    From the formula (\ref{B-C}), one can see that for a given $A$-matrix,
\begin{itemize}
\item[$\bullet$] if a column of $A$ has only zero elements, then  the exponential term $E_m$ with $m$ being
the column index never appear in the $\tau$-function, and
\item[$\bullet$] if a row of $A$ has only the pivot as non-zero element, then $E_n$ with $n$ being
the row index can be factored out from the $\tau$-function, i.e. $E_n$ has no contribution to the solution.
\end{itemize}
We then say that an $A$-matrix with no such cases  is {\it irreducible}, because the $\tau$-function with reducible matrix can be obtained by a matrix with smaller size.

We are also interested in non-singular solutions. Since the solution is given by $u=2\partial_x^2(\ln\tau)$,
the non-singular solutions are obtained by imposing the non-negativity condition on the minors,
\begin{equation}\label{NNC}
\xi(j_1,j_2,\ldots,j_N)\ge 0,\qquad {\rm for~all}\quad 1\le j_1<j_2<\cdots<j_N\le M.
\end{equation}
This condition is not only sufficient but also necessary for the non-singularity of the solution. We call a matrix having
the condition (\ref{NNC}) {\it totally non-negative} matrix.

\subsection{Line-soliton solution and the notations}
\begin{figure}[t]
\begin{centering}
\includegraphics[scale=0.8]{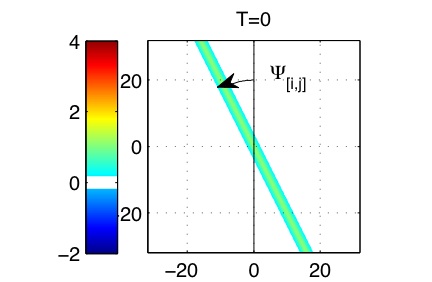}\hskip 0.8cm
\raisebox{0.5cm}{\includegraphics[height=3cm]{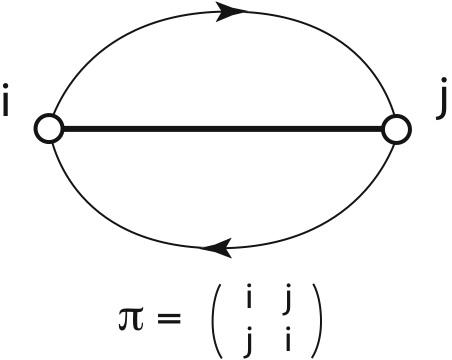}}
\par\end{centering}
\caption{One line-soliton solution of $[i,j]$-type and the corresponding chord diagram. The amplitude $A_{[i,j]}$ and the angle $\Psi_{[i,j]}$
are given by $A_{[i,j]}=\frac{1}{2}(k_i-k_j)^2$ and $\tan\Psi_{[i,j]}=k_i+k_j$.
The upper oriented chord represents
the part of $[i,j]$-soliton for $y\gg 0$ and the lower one for $y\ll 0$.
Here $i,j$ represent not just $k_i,k_j$, but also the value of those parameters,
i.e. the horizontal line is the coordinate for the $k$-parameters, and in this example, $k_1=-0.5$ and $k_2=1$.}
\label{fig:1soliton}
\end{figure}

Let us here present the simplest solution, called one {\it line-soliton} solution,
and introduce several notations to describe the solution.
A line-soliton solution is obtained by a $\tau$-function with two exponential terms, i.e.
the case $N=1$ and $M=2$ in (\ref{NM}): With the $A$-matrix of the form $A=(1~a)$,
we have
\[
\tau=E_1+aE_2=2\sqrt{a}e^{\frac{1}{2}(\theta_1+\theta_2)}\cosh\frac{1}{2}(\theta_1-\theta_2-\ln a)\,,
\]
The parameter $a$  in the $A$-matrix must be $a\ge 0$ for a {\it non-singular} solution
(i.e. totally non-negative $A$-matrix),
and it determines the location of the soliton solution.
Since $a=0$ leads to a trivial solution, we consider only $a>0$ (i.e. irreducible $A$-matrix) .
Then the solution $u=2{\partial_x^2}(\ln\tau)$ gives
\[
u=\frac{1}{2}(k_1-k_2)^2\sech^2\frac{1}{2}(\theta_1-\theta_2-\ln a).
\]
Thus the solution is located along the line $\theta_1-\theta_2=\ln a$.
We here emphasize that the line-soliton appears at the boundary of
two regions, in each of which either $E_1$ or $E_2$ becomes the dominant exponential term,
 and because of this we also call this soliton
$[1,2]$-soliton solution (or soliton of $[1,2]$-type). In general, the line-soliton solution of $[i,j]$-type
with $i<j$ has the following structure (sometimes we consider only locally),
\begin{equation}\label{1solitontheta}
u=A_{[i,j]}\sech^2\frac{1}{2}\left({\bf K}_{[i,j]}\cdot {\bf x}-\Omega_{[i,j]}t+\Theta_{[i,j]}\right)
\end{equation}
with some constant $\Theta_{[i,j]}$.
The amplitude $A_{[i,j]}$, the wave-vector ${\bf K}_{[i,j]}$ and the frequency $\Omega_{[i,j]}$ are defined by
\begin{align*}
A_{[i,j]}&=\frac{1}{2}(k_j-k_i)^2\\
{\bf K}_{[i,j]}&=\left(k_j-k_i, k_j^2-k_i^2\right)=(k_j-k_i)\left(1,k_i+k_j\right),\\
\Omega_{[i,j]}&=k_j^3-k_i^3=(k_j-k_i)(k_i^2+k_ik_j+k_j^2).
\end{align*}
The direction of the wave-vector ${\bf K}_{[i,j]}=(K_{[i,j]}^x,K_{[i,j]}^y)$ is measured in the counterclockwise from the $x$-axis, and
it is given by
\[
\frac{K^y_{[i,j]}}{K^x_{[i,j]}}=\tan\Psi_{[i,j]}=k_i+k_j.
\]
Notice that $\Psi_{[i,j]}$ gives the angle between the line ${\bf K}_{[i,j]}\cdot {\bf x} =const$ and the $y$-axis (See Figure \ref{fig:1soliton}). Then a line-soliton (\ref{1solitontheta}) can be written in the form with three parameters $A_{[i,j]},\Psi_{[i,j]}$ and $x^0_{[i,j]}$,
\begin{equation}\label{Onesoliton}
u=A_{[i,j]}\sech^2\sqrt{\frac{A_{[i,j]}}{2}} \left(x+y\tan\Psi_{[i,j]}-C_{[i,j]}t-x^0_{[i,j]}\right),
\end{equation}
with $C_{[i,j]}=k_i^2+k_ik_j+k_j^2=\frac{1}{2}A_{[i,j]}+\frac{3}{4}\tan^2\Psi_{[i,j]}$.
For the parameter $x_{[i,j]}^0$ giving the location of the line-soliton, we also use the notation,
\begin{equation}\label{xshift}
x_{[i,j]}^0=-\frac{1}{k_j-k_i}\Theta_{[i,j]},
\end{equation}
with $\Theta_{[i,j]}$ in (\ref{1solitontheta}) which is determined
by the $A$-matrix and the $k$-parameters.
For multi-soliton solutions, one can only define the location of each $[i,j]$-soliton using the asymptotic position
in the $xy$-plane either $x\gg0$ or $x\ll0$
(we mainly consider the cases where the solitons are not parallel to the $y$-axis),
and we use the notation $x^+_{[i,j]}$ (or $x^{-}_{[i,j]}$) which describes the {\it $x$-intercept}
of the line determined by the wave crest of $[i,j]$-soliton in the region $x\gg 0$ (or $x\ll 0$) at $t=0$. In Figure
\ref{fig:1soliton}, we illustrate an example of one line-soliton
solution. In the right panel of this figure, we show a {\it chord
diagram} which represents this soliton solution. Here the chord
diagram indicates the permutation of the dominant exponential terms
$E_i$ and $E_j$ in the $\tau$-function,  that is, with the ordering
$k_i<k_j$, $E_i$ dominates in $x\ll 0$, while $E_j$ dominates in
$x\gg 0$. This representation of the line-solitons in terms of the
chord diagrams is the key concept throughout the present paper. (See
section \ref{sec:classification} below for the precise definition of
the chord diagrams.)
%%%%%%%%%%%%%%%%%%%%%%%%%%%
%%%%%%%%%%%%%%%%%%%%%%%%%

For each soliton solution of (\ref{Onesoliton}),  the wave vector
${\bf K}_{[i,j]}$ and the frequency $\Omega_{[i,j]}$ satisfy the
soliton-dispersion relation, i.e.,
\begin{equation}\label{Sdispersion}
4\Omega_{[i,j]} K_{[i,j]}^x=(K_{[i,j]}^x)^4+3(K_{[i,j]}^y)^2.
\end{equation}
The soliton velocity ${\bf V}_{[i,j]}$ defined by  ${\bf K}_{[i,j]}\cdot{\bf V}_{[i,j]}=\Omega_{[i,j]}$ is given by
\[
{\bf V}_{[i,j]}=\frac{\Omega_{[i,j]}}{|{\bf K}_{[i,j]}|^2}{\bf K}_{[i,j]}=\frac{k_i^2+k_ik_j+k_j^2}{1+(k_i+k_j)^2}\,(1,\,k_i+k_j).
\]
Note in particular that $C_{[i,j]}=k_i^2+k_ik_j+k_j^2>0$, and this implies that  the $x$-component of the velocity is {\it always} positive, that is, any soliton
propagates in the positive $x$-direction. On the other hand, one should note that any small perturbation propagates
in the negative $x$-direction, that is, the $x$-component of the group velocity is
always negative. This can be seen from the dispersion relation of the KP equation
for a linear wave $\phi=\exp(i{\bf k}\cdot{\bf x}-i\omega t)$ with the wave-vector ${\bf k}=(k_x,k_y)$ and
the frequency $\omega$,
\[
\omega=-\frac{1}{4}k_x^3+\frac{3}{4}\,\frac{k_y^2}{k_x},
\]   from which the group velocity of the wave is given by
\[
{\bf v}=\nabla \omega=\left(\frac{\partial\omega}{\partial k_x},\frac{\partial \omega}{\partial k_y}\right)=
\left(-\frac{3}{4}\left(k_x^2+\frac{k_y^2}{k_x^2}\right), \frac{2}{3}\,\frac{k_y}{k_x}\right).
\]
This is similar to the case of the KdV equation, and we expect that asymptotically
soliton separates from small radiations.  Physically this implies that soliton is a {\it supersonic}
wave due to its nonlinearity (recall that the velocity of shallow water wave is proportional
to the square root of the water depth, and the KP equation in the form (\ref{kp}) describes the waves
in the moving frame with the phase velocity in the $x$-direction).

Stability of one-soliton solution was shown in
the original paper by Kadomtsev and Petviashvii \cite{KP:70}, and this may be stated as follows:
For any $\epsilon>0$ and any $r>0$, there exists $\delta>0$ so that if the initial wave $u(x,y,0):=u^0(x,y)$
satisfies
\[
\|u^0-u_{\rm exact}^0\|_{L^2(D_r^0)}<\delta,
\]
for some exact soliton solution, $u_{\rm exact}^t:=A_0\sech^2 \sqrt{\frac{A_0}{2}}(x+y\tan\Psi_0-C_0t-x_0)$
with appropriate constants $A_0,\Psi_0$ and $x_0$ (recall $C_0=\frac{1}{2}A_0+\frac{3}{4}\tan^2\Psi_0$), then the stability implies that the solution $u^t(x,y):=u(x,y,t)$ satisfies
\[
\| u^t-u^t_{\rm exact}\|_{L^2(D_r^t)}<\epsilon,\quad{\rm for}\quad t\to \infty,
\]
where $D_r^t$ is a circular disc with radius $r$ moving with the soliton, i.e.
\[
D_r^t=\{(x,y)\in \mathbb{R}^2: (x-x_0(t))^2+(y-y_0(t))^2\le r^2\},
\]
with $(x_0(t),y_0(t))$ at any point on the soliton, $x_0(t)+y_0(t)\tan\Psi_0=C_0t+x_0$.
Here $\| f\|_{L^2(D)}$ is the usual $L^2$-norm of $f(x,y)$ over a compact domain $D\subset\mathbb{R}^2$,
i.e.
\[
\|f\|_{L^2(D)}:=\left(\iint_D|f(x,y)|^2\,dxdy\right)^{\frac{1}{2}}.
\]
This stability implies a separation of the soliton from the dispersive radiations (non-soliton parts)
as in the case of the KdV soliton.
We would like to prove the similar statement for more general initial waves. However
there are several difficulties for two-dimensional stability problem in general.
In this paper, we will give a numerical study for some special cases
where the initial waves consist of two
semi-infinite line solitons with V- or X-shape.

Finally we remark that a line-soliton having the angle
$\Psi\approx \frac{\pi}{2}$ has an infinite speed, and of course it is beyond the assumption
of the quasi-two dimensionality.  We also emphasize that the structure of the solution for $y\gg 0$
can be different from that in $y\ll 0$, that is, the set of asymptotic solitons in $y\gg 0$ can be
different from that in $y\ll 0$. This difference is a consequence of the resonant interactions
among solitons as we can see throughout the paper.

\subsection{Classification Theorems}\label{sec:classification}

Now we present the main theorems obtained in \cite{CK:08,CK:09} for the classification of soliton solutions generated by the $\tau$-functions with irreducible and
totally non-negative $A$-matrices:
\begin{Theorem}\label{main1}
Let $\{e_1,\ldots, e_N\}$ be the pivot indices, and let $\{g_1,\ldots, g_{M-N}\}$ be non-pivot indices
for an $N\times M$ irreducible and totally non-negative $A$-matrix. Then
the soliton solution generated by the $\tau$-function with the $A$-matrix has the following asymptotic structure:
\begin{itemize}
\item[(a)] For $y\gg0$, there are $N$ line-solitons of $[e_n,j_n]$-type for $n=1,\ldots, N$.
\item[(b)] For $y\ll0$, there are $M-N$ line-solitons of $[i_m,g_m]$-type for $m=1,\ldots,M-N$.
\end{itemize}
Here $j_n(>e_n)$ and $i_m(<g_m)$ are determined uniquely from the $A$-matrix.
\end{Theorem}

The unique index pairings $[e_n,j_n]$ and $[i_m,g_m]$ in Theorem \ref{main1} have a combinatorial interpretation.
Let us define the pairing map $\pi:\{1,2,\ldots,M\}\to\{1,2,\ldots,M\}$ such that
\begin{equation}\label{pmap}
\left\{\begin{array}{llll}
\pi(e_n)=j_n,\qquad &n=1,\ldots,N,\\[1.0ex]
\pi(g_m)=i_m,\qquad &m=1,\ldots, M-N,
\end{array}\right.
\end{equation}
where $e_n$ and $g_m$ are respectively the pivot and non-pivot indices of the $A$-matrix.
Then we have:
\begin{Theorem}\label{permutation}
The pairing map $\pi$ is a bijection. That is, $\pi\in S_M$, where $S_M$ is the group
of permutation for the index set $\{1,2,\ldots,M\}$, i.e.
\[
\pi=\begin{pmatrix}
e_1 & \cdots & e_N & g_1 & \cdots & g_{M-N}\\
j_1 & \cdots & j_N & i_1 & \cdots & i_{M-N}
\end{pmatrix}
\]
Note in particular that the corresponding $\pi$ is the
derangement, i.e. $\pi$ has no fixed point.
\end{Theorem}

\begin{figure}[h]
\centering
\includegraphics[scale=0.43]{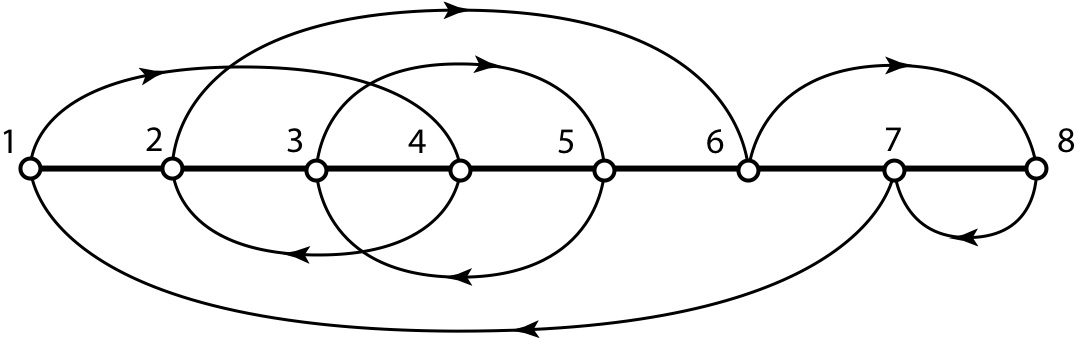}
\caption{Example of the chord diagram.  This shows $\pi=(46523817)$.
Each chord joining $i$-th and $j$-th marked points represents the line-soliton of $[i,j]$-type.}
\label{fig:chord8}
\end{figure}

Theorem \ref{permutation} shows that the pairing map $\pi$ for an $(M-N,N)$-soliton solution has $N$ {\it excedances}, i.e. $\pi(i)>i$ for $i=1,\ldots,N$,
and the excedance set is the set of pivot indices of the $A$-matrix. We represent each soliton solution
with the {\it  chord diagram} defined as follows:
\begin{itemize}
\item[(i)] There are $M$ marked points on a line, each of the point corresponds to the $k$-parameter.
\item[(ii)] On the upper side of the line, there are $N$ chords (pairings), each of them connects two points
on the line representing $\pi(e_n)=j_n$ for $n=1,\ldots,N$, i.e. the excedance $j_n>e_n$.
\item[(iii)] On the lower side of the line, there are $M-N$ chords representing $\pi(g_m)=i_m$ for
$m=1,\ldots,M-N$, i.e. the deficiency $i_m<g_m$.
\end{itemize}
In Figure \ref{fig:chord8}, we illustrate an example of the chord diagram which represents the derangement,
\[
\pi=\begin{pmatrix}
1&2&3&4&5&6&7&8\\
4&6&5&2&3&8&1&7
\end{pmatrix}\quad {\rm or~simply}\quad \pi=(46523817).
\]
The diagram then shows that the set of excedances is $\{1,2,3,6\}$, and the corresponding
soliton solution consists of the asymptotic line-solutions of $[1,4]$-, $[2,6]$-, $[3,5]$- and $[6,8]$-types in $y\gg0$
and of $[1,7]$-, $[2,4]$-, $[3,5]$- and $[7,8]$-types in $y\ll 0$.

\section{Exact solutions}\label{sec:exact}
Here we present several exact solutions generated by smaller size matrices with $N\le 3$ and $M\le 4$.
Those solutions give a fundamental structure of general solutions, and
we will show that some of those solutions appear naturally as asymptotic solutions
of the KP equation for certain classes of initial waves related to the  rogue wave generation \cite{PTLO:05, TO:07, TOM:08,F:80}.
The detailed discussions and the formulae given in this section can be found in \cite{CK:09}.

\subsection{Y-shape solitons: Resonant solutions}
We first discuss the resonant interaction among line-solitons, which is the most important
feature of the KP equation (see e.g. \cite{M:77, NR:77, KY:80}). To describe resonant solutions, let us consider the $\tau$-function with $M=3$, that is, the $\tau$-function
has three exponential terms $\{E_1,E_2,E_3\}$.
 In terms of the $\tau$-function in the form (\ref{wronskian}), those are
 given by the cases $(N=1,M=3)$ and $(N=2,M=3)$.

\begin{figure}[t]
\begin{center}
\hskip 0.2cm \includegraphics[scale=0.8]{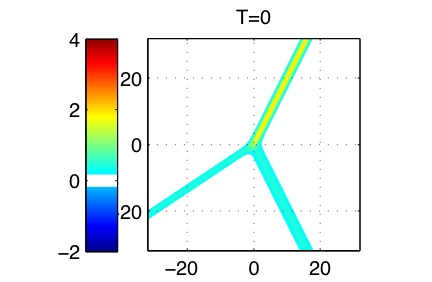} \hskip
0.9cm
\raisebox{0.5cm}{\includegraphics[height=2.5cm]{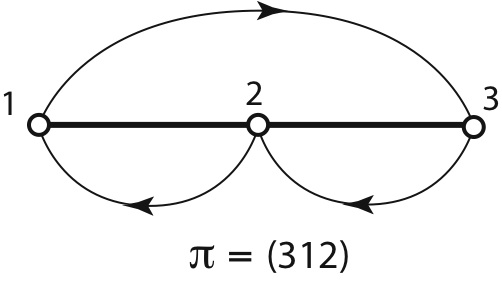}}\\[1.0ex]
\includegraphics[scale=0.8]{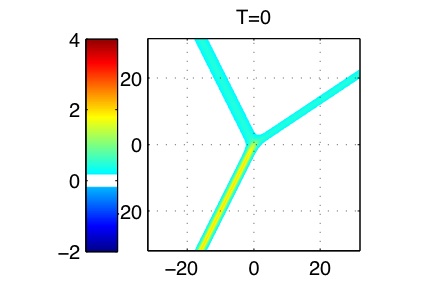} \hskip1cm
\raisebox{0.5cm}{\includegraphics[height=2.5cm]{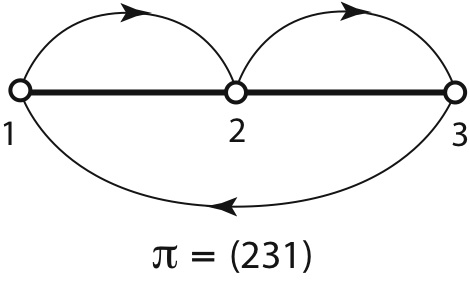}}
\caption{Examples of $(2,1)$- and $(1,2)$-soliton solutions and the
chord diagrams. The $k$-parameters are given by
$(k_1,k_2,k_3)=(-\frac{5}{4},-\frac{1}{4},\frac{3}{4})$. The right
panels are the corresponding chord diagrams. For both cases, the
parameters in the $A$-matrices are chosen so that at $t=0$ three
line-solitons meet at the origin as shown in the left
panels.}\label{fig:2}
\end{center}
\end{figure}

 Let us first study the case with $N=1$ and $M=3$, where the $A$-matrix is given by
 \[
 A=\begin{pmatrix} 1 & a & b \end{pmatrix}.
 \]
 The parameters $a,b$ in the matrix are positive constants, and the positivity implies the irreducibility and the regularity
 of the solution. The $\tau$-function is simply given by
 \[
 \tau=E_1+aE_2+bE_3.
 \]
(Note that if one of the parameters is zero, then the $\tau$-function consists only two exponential terms and
it gives one line-soliton solution, i.e. reducible case.)
With the ordering $k_1<k_2<k_3$, the solution $u=2\partial_x^2(\ln\tau)$ consists of $[1,3]$-soliton for $y\gg 0$ and
$[1,2]$- and $[2,3]$-solitons for $y\ll 0$.
Taking the balance between two exponential terms in the $\tau$-function,  one can see that
those line-solitons of $[1,3]$- and $[2,3]$-types are localized along the lines given in (\ref{Onesoliton}), i.e.
$x+y\tan\Psi_{[i,j]}-C_{[i,j]}t =x_{[i,j]}^0$ for $[i,j]=[1,3]$ and $[2,3]$ where
the locations $x_{[i,j]}^0$ are determined by the $A$-matrix (see \eqref{xshift}),
\begin{equation}\label{2-1locations}
x_{[1,3]}^0=-\frac{1}{|3,1|}\ln b,\qquad
x_{[2,3]}^0=-\frac{1}{|3,2|}\ln \frac{b}{a}.
\end{equation}
where $|i,j|:=k_i-k_j$. The shape of solution generated by $\tau=
E_1+{a}E_2+{b}E_3$ with $a=b=1$ (i.e. at $t=0$ three line-solitons
meet at the origin) is illustrated via the contour plot in the first
row of Figure \ref{fig:2}.
%%%%%%%%%%%%%%%%%%%%%%%%%%%
%%%%%%%%%%%%%%%%%%%%%%%%%%%%%%%%
%\end{Example}
This solution represents a resonant solution of three line-solitons,
and the resonant condition is
given by
\begin{align*}
{\bf K}_{[1,3]}={\bf K}_{[1,2]}+{\bf K}_{[2,3]},\qquad \Omega_{[1,3]}=\Omega_{[1,2]}+\Omega_{[2,3]},
\end{align*}
which are trivially satisfied with ${\bf K}_{[i,j]}=(k_j-k_i,k_j^2-k_i^2)$ and $\Omega_{[i,j]}=
k_j^3-k_i^3$.
The chord diagram corresponding to this soliton is
shown in the right panel of the first raw in Figure \ref{fig:2}, and it represents the permutation $ \pi=(312)$.

Let us now consider the case with $N=2$ and $M=3$: We take the $A$-matrix in the form,
\[
A=\begin{pmatrix}
1 &0 &-b \\ 0& 1 & a
\end{pmatrix}.
\]
where $a$ and $b$ are positive constants.
Then the $\tau$-function is given by
\[
\tau=E(1,2)+aE(1,3)+bE(2,3),
\]
with $E(i,j)=(k_j-k_i)E_iE_j$ for $i<j$.
In this case we have $[1,2]$- and $[2,3]$-solitons for $y\gg 0$ and
$[1,3]$-soliton for $y\ll 0$, and this solution
can be labeled by $\pi=(231)$.  Those line-solitons of $[1,2]$- and $[1,3]$-types are localized along
the lines, $x+y\tan\Psi_{[i,j]}-C_{[i,j]}t=x^0_{[i,j]}$ with
\begin{equation}\label{1-2locations}
x_{[1,2]}^0=-\frac{1}{|2,1|}\ln \left(\frac{|3,2|}{|3,1|}\frac{b}{a}\right),
\qquad x_{[1,3]}^0=-\frac{1}{|3,1|}\ln
\left(\frac{|3,2|}{|2,1|}b\right),
\end{equation}
where $|i,j|:=k_i-k_j$.
In the lower figures of Figure \ref{fig:2}, we illustrate the solution in this case.
Notice that this figure can be obtained from $(2,1)$-soliton in the upper figure by changing $(x,y)\to(-x,-y)$.  Here the parameters $a$ and $b$ in the $A$-matrix are chosen,
so that $x_{[1,2]}^0=x_{[1,3]}^0=0$, that is, all of those soliton solutions meet at the origin at $t=0$.

\subsection{$N=1,3$ and $M=4$ cases}
%%%%%%%%%%%%%%%%%%%%%%%%%%%%
\begin{figure}[t!]
\begin{centering}
\includegraphics[scale=0.60]{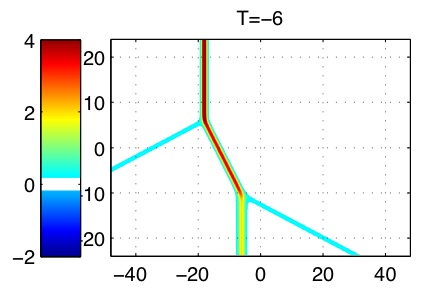}
\includegraphics[scale=0.60]{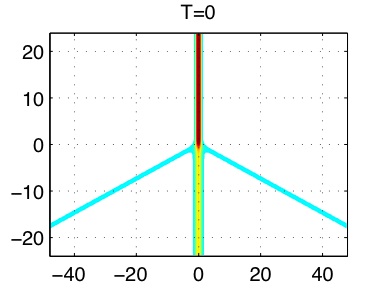}
\includegraphics[scale=0.60]{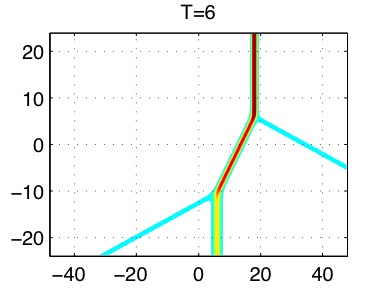}
\par\end{centering}
\begin{centering}
\includegraphics[scale=0.60]{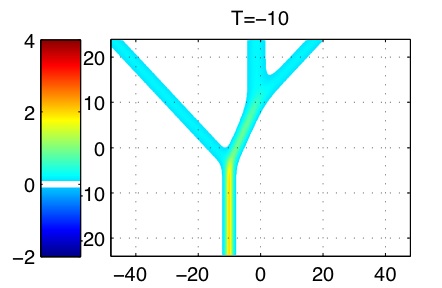}
\includegraphics[scale=0.60]{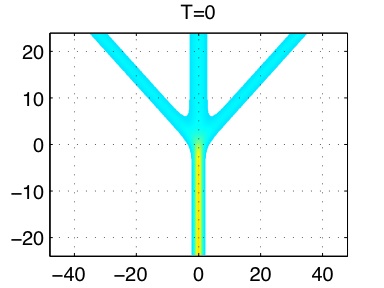}
\includegraphics[scale=0.60]{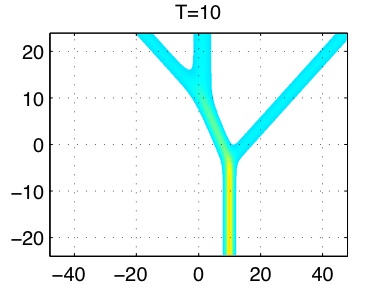}
\par\end{centering}
\caption{\label{fig:VE-1} Examples of $(3,1)$- and $(1,3)$-soliton solutions.
The solution in the upper figures are generated by the $1\times 4$ $A$-matrix with
$(k_{1},k_2,k_3,k_{4})=(-\sqrt{3},-1,1,\sqrt{3})$. The symmetry in the $k$-parameters implies that $[1,4]$- and $[2,3]$-solitons are parallel to the $y$-axis.
The lower figures are generated by the $3\times 4$ $A$-matrix with
$(k_{1},k_2,k_3,k_{4})=(-1,-\frac{1}{2\sqrt{2}},\frac{1}{2\sqrt{2}},1)$.}
\end{figure}
%%%%%%%%%%%%%%%%%%%%%%%%%%%%%%%%%%%%%%%
Let us first discuss the case with $N=1$ and $M=4$, that is, the $A$-matrix is given by
\[
A=(1~a~b~c),
\]
where $a,b$ and $c$ are positive constants. The $\tau$-function is simply written in the form
\[
\tau=E_1+aE_2+bE_3+cE_4.
\]
In this case, we have $(3,1)$-soliton solution consisting of one line-soliton of $[1,4]$-type for $y\gg 0$ and
three line-solitons of $[1,2]$-, $[2,3]$- and $[3,4]$-types for $y\ll 0$. This solution is  labeled by
$\pi=(4\,1\,2\,3)$.  The upper figures  in Figure \ref{fig:VE-1} shows the time-evolution of the solution of this type. We set $a=b=c=1$ of the $A$-matrix, so that all four line-solitons meet at the origin
at $t=0$.

  For $N=3$, any irreducible and totally non-negative $A$-matrix has the form,
\[
A=\begin{pmatrix} 1 & 0 & 0 & c\\ 0&1&0 &-b\\0&0&1&a\end{pmatrix},
\]
where $a,b$ and $c$ are positive constants. The $\tau$-function is then given by
\[
\tau=E(1,2,3)+aE(1,2,4)+bE(1,3,4)+cE(2,3,4),
\]
with $E(l,m,n)=(k_n-k_m)(k_n-k_l)(k_m-k_l)E_lE_mE_n$.  This gives $(1,3)$-soliton solution
which is dual to the case of $N=1$, that is, $[1,4]$-soliton for $y\ll0$ and $[1,2]$-, $[2,3]$- and $[3,4]$-solitons for $y\gg0$.
The corresponding label for this solution is given by $\pi=(2\,3\,4\,1)$.  The lower figures in Figure
\ref{fig:VE-1} shows the time-evolution of this type. Here we set $a,b,c$ of the $A$-matrix as
\[
a=\frac{|1,2,3|}{|1,2,4|},\quad b=\frac{|1,2,3|}{|1,3,4|},\quad c=\frac{|1,2,3|}{|2,3,4|},
\]
where $|l,m,n|:=|(k_n-k_m)(k_n-k_l)(k_m-k_l)|$, so that all four solitons meet at the origin at $t=0$.

\subsection{$N=2$ and $M=4$ cases}\label{sec:exact2-4}
\begin{figure}[t!]
\centering
\includegraphics[scale=0.52]{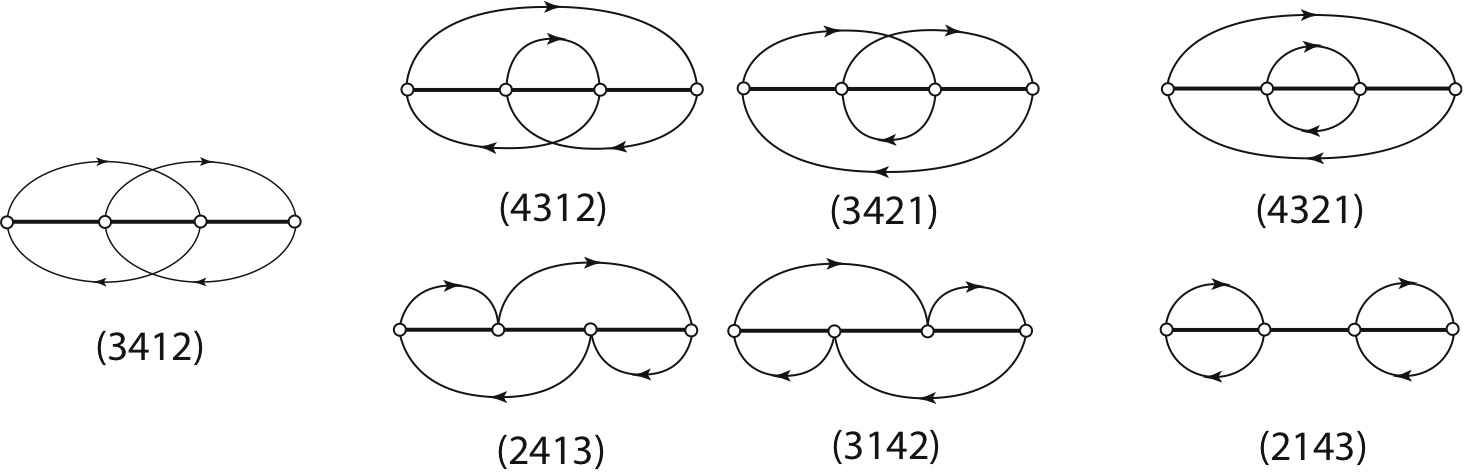}
\caption{The chord diagrams for seven different types
of $(2,2)$-soliton solutions. Each diagram parametrizes a unique cell in the totally non-negative Grassmannian Gr$(2,4)$.
}
\label{fig:chords}
\end{figure}

By a direct construction of the derangements of $S_4$ with two exedances (i.e. $N=2$),
one can easily see that there are seven cases with $2\times 4$ irreducible and
totally non-negative $A$-matrices.  Then the classification theorems imply that we have a $(2,2)$-soliton solution associated to each of those $A$-matrices. We here discuss all of those $(2,2)$-soliton solutions
with the same $k$-parameters given by
 $(k_1,k_2,k_3,k_{4})=(-\frac{7}{4},-\frac{1}{4},\frac{3}{4},\frac{3}{2})$, and show how each $A$-matrix
chooses a particular set of line-solitons. In Figure \ref{fig:chords}, we illustrate the
corresponding chord diagrams for all those seven cases, from which one can find the asymptotic
line-solitons in each case.
%%%%%%%%%%%%%%%%%%%%%%%%%%%%%%%%%%%%%%%%%%
%%%%%%%%%%%%%%%%%%%%%%%%%%%%%%%%%%%

\subsubsection{The case $\pi=(3412)$ }
\begin{figure}[h]
\begin{centering}
\includegraphics[scale=0.6]{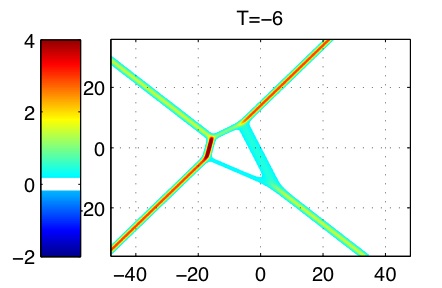}\hskip -2mm
\includegraphics[scale=0.6]{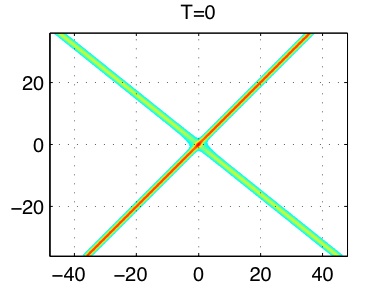}\hskip -3mm
\includegraphics[scale=0.6]{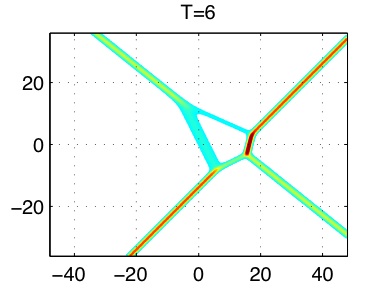}
\par\end{centering}
\caption{\label{fig:3412} Example of $(3412)$-type soliton solution (T-type). The asymptotic solitons are $[1,3]$- and $[2,4]$-types for both $y\to\pm\infty$. The intermediate solitons forming a box-shape pattern are given by
$[1,2]$-,$[1,4]$-, $[2,3]$- and $[3,4]$-solitons.}
\end{figure}

From the chord diagram in Figure \ref{fig:chords}, one can see that the asymptotic line-solitons are given by
$[1,3]$- and $ [2,4] $-types for both $y\to\pm\infty$.  The solution in this case corresponds to
the top cell of Gr$(2,4)$, and
the $A$-matrix is given by
\[
A=\begin{pmatrix} 1&0&-c&-d\\0&1&a&b\end{pmatrix}\,,
\]
where $a,b,c,d>0$ are free parameters with $D:=ad-bc>0$. This is the generic solution
on the maximal dimensional cell, and it is called T-type (after \cite{K:04}).
The most important feature of this solution is
the generation of four intermediate solitons forming a box at the intersection point.
Those intermediate solitons are identified as $[1,2]$-, $[2,3]$-, $[1,4]$- and $[3,4]$-solitons, and
they may be obtained by {\it cutting} the chord diagram of {{$(3412)$}}-type at the crossing
points. For example, if we cut the chords of $[1,3]$ and $[2,4]$ at the crossing point,
we obtain either pair of $\{[1,2],[3,4]\}$ or $\{[1,4],[2,3]\}$. The first pair appears in
the lower and upper edge of the box, and the second pair in the right and left edges.
Figure \ref{fig:3412} shows the time-evolution of the solution of this type.
The parameters $a,b,c$ and $d$ in the $A$-matrix determine the locations of
solitons, their phase shifts and the on-set of the box-shaped interaction pattern:
Those are given by
\begin{equation}\label{TA}
b=\frac{|2,1|}{|4,1|}se^{\Theta^+_{[1,3]}}, \quad
c=\frac{|2,1|}{|3,2|}se^{\Theta^+_{[2,4]}},\quad
\frac{a}{d}=\frac{|4,2|}{|3,1|}r,\quad D=\frac{|2,1|}{|4,3|}s,
\end{equation}
where $s$ is for the phase shift, $\Theta^+_{[i,j]}$ for the location of $[i,j]$-soliton for $x\gg 0$
(recall the definition below (\ref{xshift}), i.e. $x^+_{[i,j]}=-\frac{1}{|j,i|}\Theta^+_{[i,j]}$ gives the $x$-intercept
of the line of the $[i,j]$-soliton for $x\gg 0$  at $t=0$), and
$r$ for the on-set of the box. The phase shifts for $[1,3]$- and $[2,4]$-solitons are given by
$\Delta x_{[i,j]}:=x^-_{[i,j]}-x^+_{[i,j]}=-\frac{1}{|j,i|} \Theta_{\rm T}$ with
\begin{equation}\label{Tshift}
\Theta_{\rm T}=\ln\left(\frac{|1,4||2,3|}{|1,2||3,4|}~\frac{bc}{D}\right).
\end{equation}
In Figure \ref{fig:3412},  we have chosen those parameters as
$s=1, \Theta_{[1,3]}^+=\Theta^+_{[2,4]}=0$ and $r=1$,
so that at $t=0$ the solution forms an X-shape without phase shifts and opening of a box at the origin. In Figure \ref{fig:3412}, the amplitudes
are $A_{[1,3]}={{\frac{25}{8}}}$ and $A_{[2,4]}={{\frac{49}{36}}}$.
Among the soliton solutions for $N=2$ and $M=4$,  this solution is the most complicated and interesting
one. As you can see below, this solution contains all other solutions as some parts of this solution,
that is, as explained above, all six possible solitons  (i.e. $\binom{4}{2}=6$) appear in this
solution.

\subsubsection{The case $\pi=(4312)$} 
\begin{figure}[h]
\begin{centering}
\includegraphics[scale=0.6]{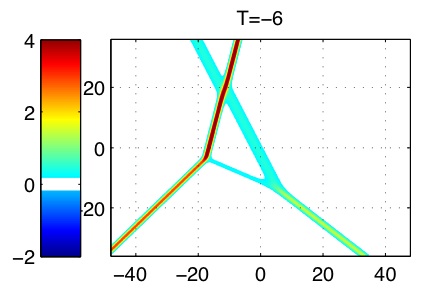}\hskip -2mm
\includegraphics[scale=0.6]{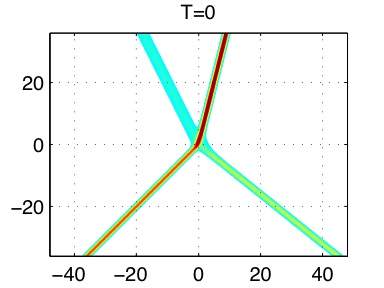}\hskip -3mm
\includegraphics[scale=0.6]{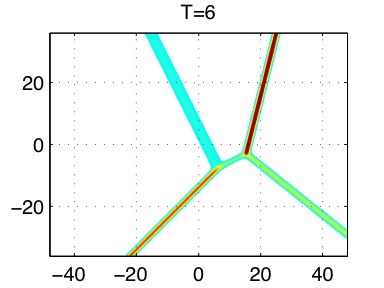}
\par\end{centering}
\caption{\label{fig:4312} Example of $(4312)$-type soliton solution.
The asymptotic solitons are $[1,4]$- and $[2,3]$-types for $y\gg 0$
and $[1,3]$- and $[2,4]$-types for $y\ll 0$. The intermediate
solitons are $[3,4]$-type for $t=-6$ and $[1,2]$-type for $t=6$.}
\end{figure}

From the chord diagram in Figure \ref{fig:chords}, one can see that the asymptotic line-solitons are given by $[1,4]$- and $[2,3]$-solitons for $y\gg0$ and
$[1,3]$- and $[2,4]$-solitons in $y\ll0$.
The $A$-matrix in this case is given by
\[
A=\begin{pmatrix} 1&0&-b&-c\\0&1&a&0\end{pmatrix}\,,
\]
where $a,b,c>0$ are free parameters.  Figure \ref{fig:4312} illustrates an example
of this solution. 
The parameters in the $A$-matrix determine the locations of the line-solitons in the
following form,
\[
a=\frac{|2,1|}{|3,1|}e^{\Theta^-_{[2,3]}},\quad
b=\frac{|2,1|}{|3,2|}e^{\Theta^-_{[1,3]}} \quad {\rm and}\quad c=
\frac{|3,1|}{|4,3|}e^{\Theta^+_{[1,4]}}.
\]
The $\Theta^{\pm}_{[ij]}$ represent the locations of the $[i,j]$-soliton for $x\gg 0$ and $x\ll 0$, and the three line-solitons determine
the location of the other one.
We here take the parameters
$\Theta^-_{[1,3]}=\Theta^+_{[1,4]}=\Theta^-_{[2,3]}=0$, so that all those
four solitons meet at the origin at $t=0$.  Notice here that the
pattern in the lower part ($y<0$) of the solution is the same as
that of the previous one, T-type. This can be also seen by comparing
the chord diagrams of those two cases.

\subsubsection{The case $\pi=(3421)$} 
\begin{figure}[h]
\begin{centering}
\includegraphics[scale=0.6]{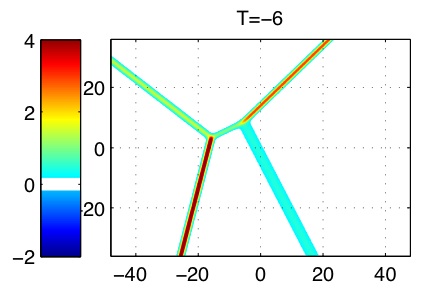}\hskip -2mm
\includegraphics[scale=0.6]{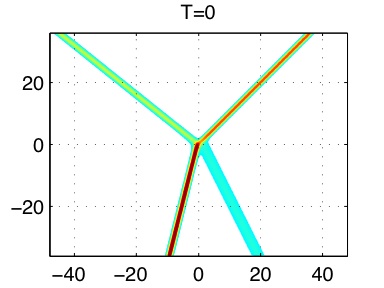}\hskip -3mm
\includegraphics[scale=0.6]{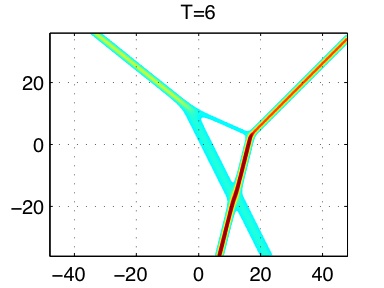}
\par\end{centering}
\caption{\label{fig:3421} Example of $(3421)$-type soliton solution.
The asymptotic solitons are given by $[1,3]$-and $[2,4]$-types for $y\gg 0$ and $[1,4]$- and $[2,3]$-types for $y\ll 0$. This solution is {\it dual} to the previous one of $(4312)$-type, i.e. the duality is given by $(x,y,t)\to (-x,-y,-t)$.}
\end{figure}

The asymptotic line-solitons in this case are given by
$[1,3]$- and $[2,4]$-solitons for $y\gg 0$ and  $[1,4]$- and $[2,3]$-solitons in $y\ll 0$.
The $A$-matrix has the form,
\[
A=\begin{pmatrix} 1&0&0&-c\\0&1&a&b\end{pmatrix}\,,
\]
where $a,b,c>0$ are free parameters.  Figure \ref{fig:3421} illustrates the evolution of the solution of this type.
The parameters in the $A$-matrix are related to the locations of the line-solitons,
\[
a=\frac{|2,1|}{|3,1|}e^{\Theta^+_{[2,3]}},\quad
b=\frac{|2,1|}{|4,1|}e^{\Theta^-_{[2,4]}}\quad {\rm and}\quad
c=\frac{|2,1|}{|4,2|}e^{\Theta^-_{[1,4]}}.
\]
This solution can be considered as a dual of the previous case (b), that is,
two sets of line-solitons for $y\gg 0$ and $y\ll 0$ are exchanged.
 Here we take
$\Theta^{\pm}_{[i,j]}=0$ for those pairs $(i,j)=(1,4), (2,3)$ and $(2,4)$, hence all those four solitons
meet at the origin at $t=0$. Note here that now the upper pattern of the solution is the same as
that of the T-type.  So if one cuts Figures \ref{fig:4312} and \ref{fig:3421} along the $x$-axis and glues the lower half of Figure \ref{fig:4312} together with the upper half of Figure \ref{fig:3421},
we can obtain Figure \ref{fig:3412} (T-type).  This can be observed from the corresponding
chord diagrams.

One should also note that this solution is ``dual'' to the previous one in the sense that the patterns of those
solutions are symmetric with respect to the change $(x,y,t)\leftrightarrow (-x,-y,-t)$.

\subsubsection{The case $\pi=(2413)$} 
\begin{figure}[h]
\begin{centering}
\includegraphics[scale=0.6]{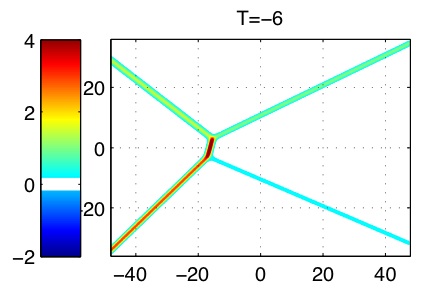}\hskip -2mm
\includegraphics[scale=0.6]{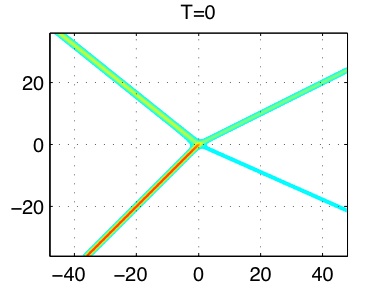}\hskip -3mm
\includegraphics[scale=0.6]{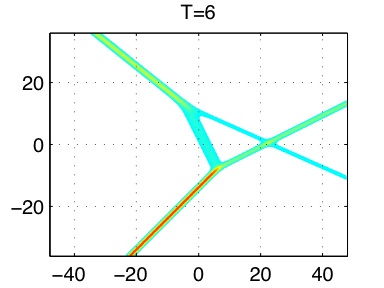}
\par\end{centering}
\caption{\label{fig:2413} Example of $(2413)$-type soliton solution. The asymptotic solitons are
$[1,2]$- and $[2,4]$-types for $y\gg0$ and $[1,3]$- and $[3,4]$-types for $y\ll0$.
The intermediate solitons are $[1,4]$-type in $t={{-6}}$ and $[2,3]$-type in $t={{6}}$.}
\end{figure}

The asymptotic line-solitons are given by
$[1,2]$- and $[2,4]$-solitons for $y\gg 0$  and  $[1,3]$- and $[3,4]$-solitons for $y\ll0$.
The $A$-matrix is given by
\[
A=\begin{pmatrix} 1&0&-c&-d\\0&1&a&b\end{pmatrix}\,,
\]
where $a,b,c,d>0$ with $ad-bc=0$.  Those parameters are related to
the locations of the line-solitons,
\[
b=\frac{|2,1|}{|4,1|}e^{\Theta^-_{[2,4]}},\quad
c=\frac{|2,1|}{|3,2|}e^{\Theta^-_{[1,3]}}\quad{\rm and}\quad
\frac{d}{b}=\frac{|4,1|}{|4,2|}e^{\Theta^+_{[1,2]}},
\]
 with $a=bc/d$.
Figure \ref{fig:2413} illustrates the evolution of the solution of this type.
For the case that all the solitons meet at the origin at $t=0$, we set all $\Theta_{[i,j]}$ to be zero.
Notice that the nonlinear interaction generates the intermediate soliton of $[1,4]$-type in $t<0$. This has the maximum amplitude among the (local) solitons appearing in the solution.
It is interesting to note that the maximum amplitude soliton might be expected from
the chord diagram, that is, this soliton is generated by the resonant interactions of
$[1,2]+[2,4]=[1,4]$ for $y>0$ and $[1,3]+[3,4]=[1,4]$ for $y<0$.  One should also note that the left patterns in Figure \ref{fig:2413} including $[1,4]$-soliton for $t<0$ and $[2,3]$-soliton for $t>0$ are the same as those
in Figure \ref{fig:3412} (T-type).

\subsubsection{The case $\pi=(3142)$} 
\begin{figure}[h]
\begin{centering}
\includegraphics[scale=0.6]{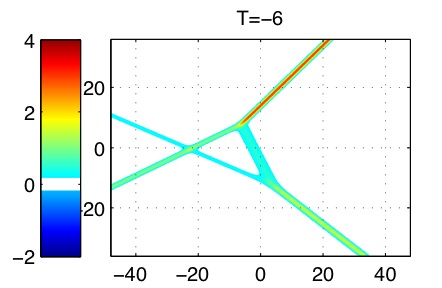}\hskip -2mm
\includegraphics[scale=0.6]{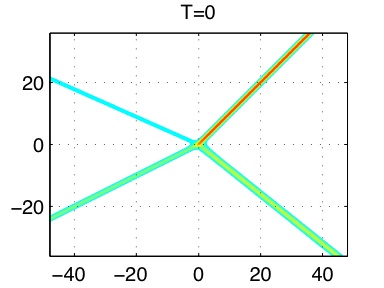}\hskip -3mm
\includegraphics[scale=0.6]{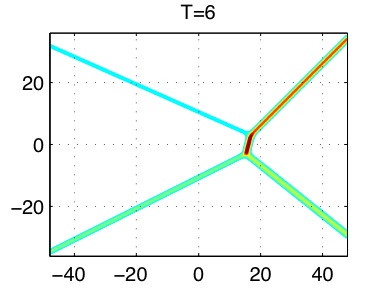}
\par\end{centering}
\caption{\label{fig:3142} Example of $(3142)$-type soliton solution. The asymptotic solutions
are $[1,3]$- and $[3,4]$-types for $y\gg0$ and $[1,2]$- and $[2,4]$-types for {{$y\ll 0$}}.
Notice the duality of this with $(2413)$-type in the previous example.}
\end{figure}

The asymptotic line-solitons are given by
$[1,3]$- and $[3,4]$-solitons in $y\gg 0$ and
 $[1,2]$- and $[2,4]$-solitons in $y\ll 0$.
The $A$-matrix has the form,
\[
A=\begin{pmatrix} 1&a&0&-c\\0&0&1&b\end{pmatrix}\,,
\]
where $a,b,c>0$. We represent those parameters in the following form which is particularly useful
for our stability problem discussed in the next section,
\begin{equation}\label{3142A}
a=\frac{|3,1|}{|3,2|}se^{\Theta^+_{[2,4]}},\quad
b=\frac{|3,1|}{|4,1|}se^{\Theta^+_{[1,3]}}\quad {\rm and}\quad
c=\frac{|3,1|}{|4,3|}s.
\end{equation}
Here $\Theta^+_{[i,j]}$ represent the location of the $[i,j]$-soliton in $x\gg0$, and $s$ represents the location
of the other line-solitons in $x\ll 0$, that is the generation of the intermediate line-solitons either $[1,4]$- or $[2,3]$-type (c.f. the case (c) in Section \ref{sec:result}).
This solution is dual to the previous one of $\pi=(2413)$.
In fact, Figure \ref{fig:3142} illustrates the dual behavior of the evolution in the sense of
the reverse of time. Here we take $s=1$ and $\Theta^+_{[1,3]}=\Theta^+_{[2,4]}=0$,
so that all solitons meet at the origin at $t=0$.
As in the previous case, the large amplitude soliton of $[1,4]$-type is generated by the
resonant interaction. In particular, this solution can be used to explain the Mach reflection
observed in shallow water \cite{M:77, F:80, CK:09}.

\subsubsection{The case  $\pi=(4321)$} 
\begin{figure}[h]
\begin{centering}
\includegraphics[scale=0.6]{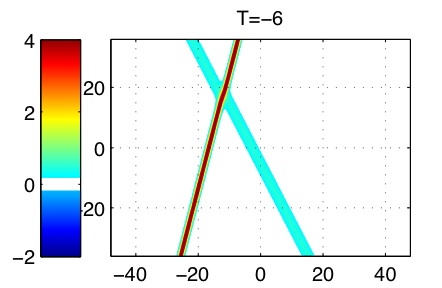}\hskip -2mm
\includegraphics[scale=0.6]{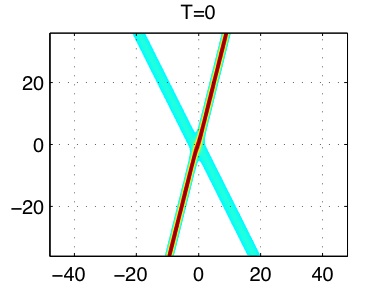}\hskip -3mm
\includegraphics[scale=0.6]{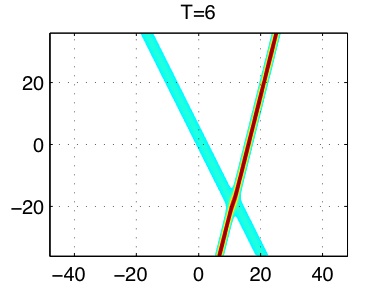}
\par\end{centering}
\caption{\label{fig:4321} Example of $(4321)$-type soliton solution (P-type).
The asymptotic solitons are $[1,4]$- and $[2,3]$-types for both $y\to\pm\infty$.
The pattern of the solution is stationary, and the interaction generates a negative phase {{shift}}
of the solitons.
}\end{figure}

The asymptotic line-solitons are
$[1,4]$- and $ [2,3] $-types for both $y\to\pm\infty$. Since those solitons can be placed
both in nearly parallel to the $y$-axis,
this type of solutions fits better in the {\it physical}
assumption for the derivation of the KP equation, i.e. a quasi-two dimensionality.
This is then referred to as P-type (after \cite{K:04}). The situation is similar to the KdV case, for example,
two solitons must have different amplitudes, $A_{[1,4]}>A_{[2,3]}$.
The $A$-matrix is given by
\[
A=\begin{pmatrix} 1&0&0&-b\\0&1&a&0\end{pmatrix}\,.
\]
The parameters are expressed by
\begin{equation}\label{PA}
a=\frac{|4,2|}{|4,3|}e^{\Theta^+_{[2,3]}}\quad {\rm and}\quad
b=\frac{|3,1|}{|4,3|}e^{\Theta^+_{[1,4]}}.
\end{equation}
where $\Theta^+_{[i,j]}$ represents the location of the $[i,j]$-soliton in $x>0$ (the wave front) (notice the phase shifts for
the solitons).
Figure \ref{fig:4321} illustrates the evolution of the solution of this type.
We here set $\Theta^+_{[i,j]}=0$, and hence the solitons in $x>0$ (the wave front) meet at the
origin at $t=0$.
Note here that there is a phase shift due to the interaction, and
the shift is in the negative direction due to the repulsive force in the interaction
similar to the case of KdV solitons. The phase shifts are given by $\Delta x_{[i,j]}=-\frac{1}{|j,i|}\Theta_{\rm P}$
with
\begin{equation}\label{Pshift}
\Theta_{\rm P}=\ln\left(\frac{|4,2||3,1|}{|2,1||4,3|}\right)>0.
\end{equation}
This negative phase shift is due to the {\it repulsive} force in the interaction as in the case of the KdV solitons.
One should also note that  the interaction point moves in the negative $y$-direction,
because the larger soliton of $[1,4]$-type propagates faster than the other one.
Those $[1,4]$- and $[2,3]$-solitons only appear as the intermediate solitons in the solution
of T-type, that is, they form a part of the box in Figure \ref{fig:3412}.

\subsubsection{The case $\pi=(2143)$ }
\begin{figure}[h]
\begin{centering}
\includegraphics[scale=0.6]{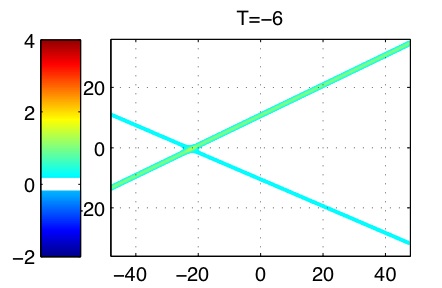}\hskip -2mm
\includegraphics[scale=0.6]{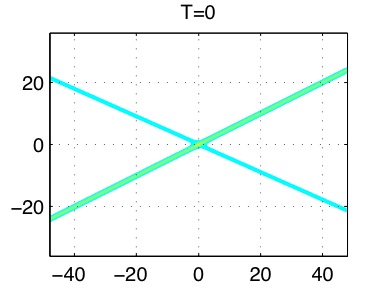}\hskip -3mm
\includegraphics[scale=0.6]{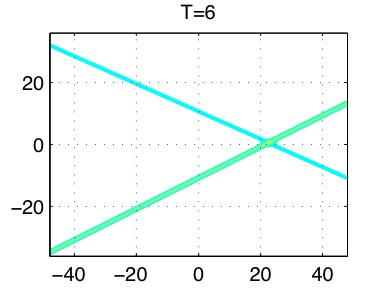}
\par\end{centering}
\caption{\label{fig:2143} Example of $(2143)$-type soliton solution (O-type).
The asymptotic solitons are $[1,2]$- and $[3,4]$-types for both $y\to\pm\infty$.
The pattern of the solution is stationary, and has a positive phase shift in the solitons.}
\end{figure}

The asymptotic line-solitons are
$[1,2]$- and $ [3,4] $-types for both $y\to\pm\infty$. This solution
was used {\it originally} to describe two soliton solution, and it is called O-type (after \cite{K:04}). Interaction properties for solitons with equal amplitude has been discussed using this solution. However this solution becomes singular,
when those solitons are almost parallel to each other and close to the $y$-axis, contrary to the assumption of the
quasi-two dimensionality for the KP equation \cite{M:77}. The $A$-matrix is given by
\[
A=\begin{pmatrix} 1&a&0&0\\0&0&1&b\end{pmatrix}\,.
\]
Notice that this A-matrix is the limit of that for $(3142)$-type with $c\to 0$ and $k_2\to k_3$.
We will further discuss this issue in the next section when we study the initial value problem with
V-shape initial wave.  The parameters $a$ and $b$ are expressed by
 \begin{equation}\label{OA}
 a ={{\frac{|4,1|}{|4,2|}}}e^{\Theta^+_{[1,2]}}\quad {\rm and}\quad b = {{\frac{|3,2|}{|4,2|}}}e^{\Theta^+_{[3,4]}}.
 \end{equation}
 where $\Theta^+{[i,j]}$ represents the location of the wave front consisting of $[1,2]$- and $[3,4]$-solitons.
 Figure \ref{fig:2143} illustrates the evolution of this type solution.
Here we take  all $\Theta^+_{[i,j]}=0$, the solitons in $x>0$ (the wave front) meet at the
origin at $t=0$. The phase {{shifts}} of those solitons are positive due to
an {\it attractive} force in the interaction, and they are given by
$\Delta x_{[i,j]}=-\frac{1}{|j,i|}\Theta_{\rm O}$ with
\begin{equation}\label{Oshift}
\Theta_{\rm O}=\ln\left(\frac{|4,1||3,2|}{|3,1||4,2|}\right)<0.
\end{equation}
 When $k_2$ approaches to $k_3$, the phase shift
gets larger, and the interaction part  of the solution is getting to be close to $[1,4]$-soliton
as a result of the resonant interaction with $[1,2]$- and $[3,4]$-solitons at $k_2=k_3$.
This is the main argument of the paper \cite{M:77} about the discussion on the Mach reflection
of shallow water waves.
Note also that those $[1,2]$- and $[3,4]$-solitons appear as the intermediate solitons
forming a box of T-type as shown in Figure \ref{fig:3412}.

\section{Numerical simulations}\label{sec:numerics}
The main purpose of the numerical simulation is to study the interaction properties
of line-solitons, and we will show that the solution of the initial value problem with
certain types of initial waves asymptotically approaches to some of the exact solutions
discussed in the previous section. This implies a stability of those exact solutions under
the influence of certain deformations (notice that the deformations in our initial waves are not so small).

The initial value problem considered here is essentially an infinite
energy problem in the sense that each line-soliton in the initial
wave is supported asymptotically in either $y\gg 0$ or $y\ll 0$, and
the interactions occur only in a finite domain in the $xy$-plane. In
the numerical scheme, we consider the rectangular domain $D=\{(x,y):
|x|\le L_x,~ |y|\le L_y\}$, and each line-soliton is matched with a
KdV soliton at the boundaries $y=\pm L_y$. In this section, we
explain the details of our numerical scheme and give an error
estimate of the scheme.

\subsection{Numerical scheme}
%\textcolor{blue}{This section is new}

Using the so-called {\it window technique} \cite{S:05, TOM:08}, we first transform our non-periodic solution $u$
having non-vanishing boundary values into a function $v$ rapidly decaying near the boundaries
$y=\pm L_y$, so that we consider the function $v$ to be periodic in both $x$ and $y$.
 With the window function $W(y)$, this can be expressed by
the decomposition of the solution $u$, i.e.
\[
u=v+(1-W)u\quad {\rm with}\quad v=Wu.
\]
We take the window function $W(y)$ in the form of a super-Gaussian,
\begin{equation}\label{W}
W(y)=\exp\left(-a_n\left|\frac{y}{L_y}\right|^n\right),
\end{equation}
where $a_n$ and $n$ are positive constants (we choose $a_n=(1.111)^n\ln 10$ with $n=27$
\cite{TOM:08}).

We assume that near the boundaries $y=\pm L_y$, the solution can be described by exact line-solitons (i.e. we take $L_y$ large enough so that the interaction will not influence those parts
near boundaries up to certain finite time). We then consider the transformation in the form,
\begin{equation}\label{transformation}
u=v+(1-W)u_0,
\end{equation}
where $u_0$ consists of exact line-solitons satisfying
\[
(1-W)u_0=(1-W)u.
\]
This assumption prevents any disturbance from the boundary, and keeps the stability of the far fields
consisting of well separated line-solitons. We are interested in studying an asymptotic behavior near
the interaction point, and we do not expect to see any strong effect from the boundary region in a short time
(recall that the system has a finite group velocity for waves with non-zero $k_x$-component).

With the transformation (\ref{transformation}), we have the equation for $v$,
\begin{align}
&\partial_x\left(4\partial_t v+6v\partial_x v+\partial^3_x v\right)+3\partial_y^2v \nonumber\\
&=6(1-W)\partial_x\left(Wu_0\partial_x u_0-\partial_x(vu_0)\right)+3(2W'\partial_yu_0+W''u_0).
\label{v-equation}
\end{align}
Assuming the solution $v$ to be periodic in both $x$ and $y$ with zero boundary condition
at $|y|=L_y$, we solve (\ref{v-equation}) by using
fast Fourier transform (FFT).  For a convenience, let us rescale the domain
$D=\{(x,y):|x|\le L_x,|y|\le L_y\}$ with ${\mathcal D}:=\{(X,Y): |X|\le \pi,~|Y|\le \pi\}$, i.e.
 \[
X=\frac{\pi}{L_x}x,\quad Y=\frac{\pi}{L_y}y.
\]
Then (\ref{v-equation}) becomes
\begin{align}
\partial_{X}\left(\partial_{t}v+Pv\partial_{X}v+Q\partial_X^3v\right)+R\partial_Y^2v=F,
\label{v-scale-equation}
\end{align}
where $F$ is the right hand side of (\ref{v-equation}) with the scalings, and
 \[
P=\frac{3\pi}{2L_x},\quad Q=\frac{\pi^3}{4L_x^3},\quad\mbox{and}\quad R=\frac{3\pi L_x}{4L_y^2}.\]
Under the periodic assumption of $v$, we have
\[
v(X,Y,t)=\sum_{l=-\infty}^{\infty}\sum_{m=-\infty}^{\infty}\hat{v}(l,m,t)e^{i(lX+mY)}.
\]
Here the Fourier transformation $\hat{v}:=\mathcal{F}(v)$ is given by
\[
\hat{v}(l,m,t)=\frac{1}{(2\pi)^2}\int_{-\pi}^{\pi}\int_{-\pi}^{\pi}v(X,Y,t)e^{-i(lX+mY)}dXdY,\]
and the equation becomes
\[
\partial_{t}\hat{v}+i\left(\frac{Rm^{2}}{2l}-Ql^{3}\right)\hat{v}+{{il\frac{P}{2}}}\mathcal{N}(\hat{v})=\frac{1}{il}\mathcal{F}(F),\]
with
\[\mathcal{N}(\hat{v})=\mathcal{F}(v^{2}).
\]
Let $c=\frac{Rm^{2}}{2l}-Ql^{3}$ and $\hat{V}=e^{ict}\hat{v}$. Then we have
\begin{align}
\partial_t\hat{V}+{{il\frac{P}{2}}}e^{ict}\mathcal{N}(\hat{v})=\frac{1}{il}e^{ict}\mathcal{F}(F).
\label{v-fft-equation}
\end{align}
We write this equation in the form of time evolution,
\[
\partial_t\hat{V}={G}(t,\hat{V}).
\]
and use the 4-th order Runge-Kutta (RK4) method to solve the equation:
Namely denoting $t_n=n \Delta t$ and
\begin{align*}
K_{1} &= G\left(t_{n},\hat{V}_{n}\right),\quad
K_{2} = G\left(t_{n}+\frac{1}{2}\Delta t,\hat{V}_{n}+\frac{K_{1}}{2}\Delta t\right),\\
K_{3} &=G\left(t_{n}+\frac{1}{2}\Delta t,\hat{V}_{n}+\frac{K_{2}}{2}\Delta t\right),\quad
K_{4} =G\left(t_{n}+\Delta t,\hat{V}_{n}+\frac{K_{3}}{2}\Delta t\right),
\end{align*}
 we have
\[
\hat{V}_{n+1} = \hat{V}_{n}+\frac{1}{6}\left[K_{1}+2(K_{2}+K_{3})+K_{4}\right]\Delta t.
\]
We use a pseudo-spectral method to evaluate nonlinear term
$\mathcal{N}(\hat{v})$ in \eqref{v-fft-equation} and the 2nd order
central difference to approximate three derivative terms in the $F$
on the right hand side of \eqref{v-scale-equation} for simplicity.
This is because $u_0$ in \eqref{transformation} is a non-periodic
function and the estimation of derivatives from a Fourier spectral
method still introduces errors at the computational boundary. A
better choice is to provide the exact closed forms for the $x$-and
$y$-derivatives of $u_0$. Since some of the exact solutions have
rather complicated formulae, we simply use 2nd order central
difference approximation. Thus the numerical solutions are expected
to be of 2nd order accuracy when there are enough Fourier modes in
space and the solutions are vanished in large $x$.

%%%%%%%%%%%%%%%%%%%%%%%%%%%%%%%%%%%
\begin{figure}[t!]
\begin{centering}
\includegraphics[scale=1]{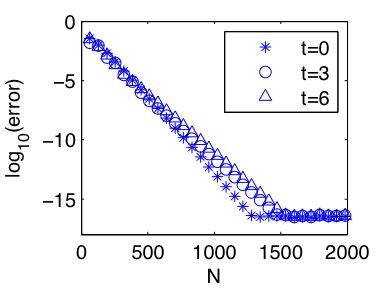}
\par\end{centering}
\caption{\label{fig:2Daccu} Numerical accuracy test on T-type
solution shown in Fig. \ref{fig:3412}. The graph shows the estimate
of max$|\mathcal{F}(v^t)-\mathcal{F}(v^t_{\rm exact})|$ as a
function of the number of Fourie modes $N \times N/2$ for $t=0,3$
and $6$. The error is decreasing exponentially with respect to $N$
until it saturates at round of error $\approx 10^{-16}$ for $N
\approx 1280, 1472$ and $1536$ for $t= 0,3$ and $6$, respectively.
The numerical domain is $D=[-64,64]\times[-32,32]$. }
\end{figure}
%%%%%%%%%%%%%%%%%%%%%%%%%%%%%%%%%%%%%%%%%%%%%%
In order to verify the accuracy of the method, we provide two tests. The first test intends to answer the question on how many Fourier modes are needed to ensure machine accuracy in space for the solution $v$ of \eqref{v-equation}. The second test shows the overall algorithm is 2nd order accurate. In Figure \ref{fig:2Daccu}, we show the maximum error in the Fourier modes of the T-type solution in the sense that we estimate max$|\mathcal{F}(v)-\mathcal{F}(v_{\rm exact})|$ with respect to the number of Fourier modes $N$ in both $x$- and $y$-directions for the case shown in  Figure \ref{fig:3412}. On the computational domain $D=[-64,64]\times[-32,32]$, machine accuracy $\approx 10^{-16}$ is reached for $N \approx 1280, 1472$ and $1536$ for $t = 0,3$ and $6$, respectively. Since the T-type solution develops a box at later time, it requires more Fourier modes to well represent the solution. From this test with $N=1536$, we find that we need roughly $\frac{1536}{2\times 64}=12$ Fourier modes per 1 unit in $x$ to reach $10^{-16}$ accuracy. Notice that it might take longer time to run such a fine mesh for a very large computational domain. If we choose to use $8$ Fourier modes per unit length in both $x$- and $y$-directions, the spacial error is within $10^{-10}$ and it can also speed up the computation. In Table \ref{Tab: 4th RK},
we summarize the error of the solution given by $\|  u^t - u_{\rm exact}^t\|_{L^2(D)}/ \|u_{\rm exact}^t\|_{L^2(D)}$ for two different time steps $\Delta t = 0.01$ and $\Delta t = 5\times10^{-3}$ on $D=[-128,128]\times[-16,16]$ with $2048\times 256$ Fourier modes. It can be clearly seen that the order of convergence is 2 because the error decreases by a factor around 4 when the time step is halved. If we take $\Delta t = 5\times 10^{-3}$ and keep running the simulation up to $t = 12$, the magnitude of the error is always within $2.5 \times 10^{-2}$. In the following simulations, the errors of the solutions are at this magnitude or smaller.
\begin{table}[t!]
\begin{centering}
\begin{tabular}{|c|c|c|c|}
\hline
 & $\Delta t=10^{-2}$ & $\Delta t=5\times10^{-3}$ & order\tabularnewline
\hline
\hline
$t=1$ & $1.046\times10^{-2}$ & $2.645\times10^{-3}$ & $1.98$\tabularnewline
\hline
$t=2$ & $1.893\times10^{-2}$ & $4.353\times10^{-3}$ & $2.12$\tabularnewline
\hline
$t=3$ & $2.507\times10^{-2}$ & $5.986\times10^{-3}$ & $2.07$\tabularnewline
\hline
$t=4$ & $2.992\times10^{-2}$ & $7.732\times10^{-3}$ & $1.95$\tabularnewline
\hline
$t=5$ & $3.523\times10^{-2}$ & $9.675\times10^{-3}$ & $1.86$\tabularnewline
\hline
$t=6$ & $4.211\times10^{-2}$ & $1.194\times10^{-2}$ & $1.82$\tabularnewline
\hline
\end{tabular}
\caption{\label{Tab: 4th RK}The convergence test on the domain $D=[-128,128]\times[-16,16]$ with $2048\times256$ Fourier modes. The graph shows the error $\|u^t-u^t_{\rm exact}\|_{L^2(D)}/ \|u^t_{\rm exact}\|_{L^2(D)}$ for
the time steps $\Delta t=0.01$ and $5\times 10^{-3}$. This indicates that our scheme is of 2nd order accurate in time. }
\par\end{centering}
\end{table}
%%%%%%%%%%%%%%%%%%%%%%%%%%%%%%%%%%%%%%%%%%%%

\section{Numerical results}\label{sec:result}
Now we present the results of the direct numerical simulation of the
KP equation with the following two types of initial waves consisting
of two line-solitons. In particular, we fix the amplitude of one of
the line-solitons, and place these line-solitons in the symmetric
shape with respect to the $x$-axis. This then implies that each
profile of the initial wave is determined by two parameters $A_0$
(the amplitude of other soliton) and $\Psi_0$ (the angle of the
wave-vector of the soliton):
\begin{itemize}
\item[(i)]  V-shape wave consisting of two semi-infinite line-solitons, i.e. $u(x,y,0)=u_1^0(x,y)+u_2^0(x,y)$ with
\begin{equation}\left\{\begin{array}{llll}
\displaystyle{ u^0_1(x,y) =A_0\sech^2\sqrt{\frac{A_0}{2}}(x-y\tan\Psi_0) \times H(y)}, \\[2.0ex]
u^0_2(x,y)= \displaystyle{2 \sech^2(x+y\tan\Psi_0) \times H(-y),}
\end{array}\right.
\label{Vinitial}
\end{equation}
where $H(z)$ is the unit step function,
\[
H(z)=\left\{\begin{array}{lll}
1 &\quad {\rm if}~~& z\ge 0, \\
0& \quad {\rm if}~~& z<0.
\end{array}\right.
\]
Here one should note that the initial locations $x^+_{[i,j]}$ are zero for both line-solitons, that is, they cross the origin at $t=0$. The shifts of $x^+_{[i,j]}$ observed in the simulations provide the information of the interaction property.
\item[(ii)] X-shape wave of the linear combination of two line-solitons, i.e. $u(x,y,0)=u_1^0(x,y)+u^0_2(x,y)$ with
\begin{equation}\left\{\begin{array}{llll}
\displaystyle{ u^0_1(x,y) =A_0\sech^2\sqrt{\frac{A_0}{2}}(x-y\tan\Psi_0)}, \\[2.0ex]
u^0_2(x,y)= \displaystyle{2 \sech^2(x+y\tan\Psi_0).}
\end{array}\right.
\label{Xinitial}
\end{equation}
Again note that the initial locations are $x^+_{[i,j]}=0$.
\end{itemize}
These initial waves have been considered in certain physical models
of interacting waves \cite{M:77, F:80, PTLO:05,TO:07, CK:09}. In
particular, in \cite{PTLO:05, TO:07}, the V-shape initial waves were
considered to study the generation of large amplitude waves in
shallow water.

The main result of this section is to show numerically that those
initial waves {\it converge} asymptotically to some of the exact
solutions presented in Section \ref{sec:exact}.  Here what we mean
by the {\it convergence} is as follows:  We first show that the
solution $u^t(x,y)$ at the intersection point of V- or X-shape
generates a pattern due to the resonant interaction, and we then
identify the pattern near the resonant point with that given by some
of the exact solutions $u^t_{\rm exact}(x,y)$.  The convergence is
only in a local sense, and we use the following (relative) error
estimate,
\begin{equation}\label{error}
E(t):=\|  u^t - u_{\rm exact}^t\|^2_{L^2(D^t_r)}\big/ \|u_{\rm exact}^t\|^2_{L^2(D^t_r)},
\end{equation}
where $D_r^t$ is the circular disc defined by
\[
D_r^t:=\left\{(x,y)\in\mathbb{R}^2: (x-x_0(t))^2+(y-y_0(t))^2\le r^2\right\}.
\]
Here the center $(x_0(t),y_0(t))$ is chosen as the intersection point of two lines determined from the
corresponding exact solution identified by the following steps:
\begin{itemize}
\item[1.] For a given initial wave, we first identify the type of exact solution (i.e. the chord diagram) from
the formulae, $A_{[i,j]}=\frac{1}{2}(k_i-k_j)^2$ and $\tan\Psi_{[i,j]}=k_i+k_j$.  This then gives the exact solution
with the corresponding $A$-matrix. Note here that the entries of the $A$-matrix are not determined yet.
\item[2.]  We then minimize the error function $E(t)$ of \eqref{error} at certain large time $t=T_0$ by changing the entries in the $A$-matrix, that is, we consider
\begin{equation}\label{minimization}
\underset{A{\rm -matrix}}{\rm Minimize}\,E(T_0).
\end{equation}
The minimization is achieved by adjusting the solution pattern with the exact one, that is, the pattern
determines the $A$-matrix whose entries give
 the locations of all line-solitons including the intermediate solitons in the solution.  With this $A$-matrix, the center $(x_0(t),y_0(t))$ of the circular domain $D_r^t$ is given by
 the intersection point of the lines,
 \begin{equation}\label{Vsolitons}
x+y\tan\Psi_{[i_l,j_l]}-C_{[i_l,j_l]}t=x_{[i_l,j_l]}^+\qquad {\rm for}\quad l=1,2,
\end{equation}
where the locations $x_{[i_l,j_l]}^+$ are expressed in terms of the $A$-matrix (see Section \ref{sec:exact}).
Here we have $A_{[i_1,j_1]}=A_0,~ A_{[i_2,j_2]}=2$ and $\Psi_{[i_2,j_2]}=-\Psi_{[i_1,j_1]}=\Psi_0$.
The time $T_0$ may be taken as $T_0=0$, but for most of the cases, we need $T_0$ to be sufficiently large.
This implies that the solutions have nontrivial location shifts, i.e. $x^+_{[i,j]}\ne 0$ in the solution.
 \item[3.]  We then confirm that $E(t)$ further decreases for a larger time $t>T_0$ up-to a time $t=T_1>T_0$, just before the effects
of the boundary enter the disc $D_r^t$ (those effects include the periodic condition in $x$ and a mismatch on the boundary patching).
\end{itemize}
We take the radius $r$ in $D_r^t$ large enough so that the main interaction
 area is covered for all $t<T_0$, but $D_r^t$ should be kept away  from the boundary
 to avoid any influence coming from
 the boundaries. The time $T_1>T_0$ gives an
 enough time to develop a pattern close to the corresponding exact solution, but it is also limited
 to avoid any disturbance  from the boundaries for $t<T_1$.
In this paper, we give $T_0$ and $T_1$ based on the observation of the numerical results, and
we will report the detailed analysis in a future communication.

\subsection{$V$-shape initial waves}
\begin{figure}[t!]
\centering
\includegraphics[scale=0.54]{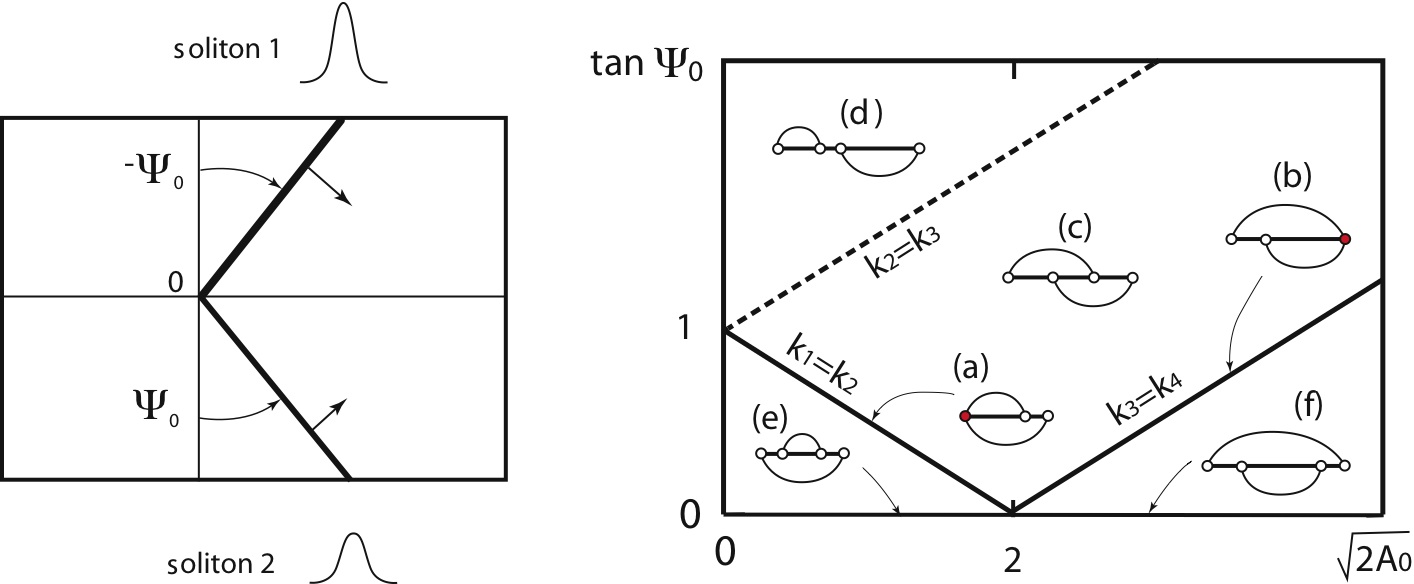}
\caption{V-shape initial waves. Each line of the V-shape is a
semi-infinite line-soliton. We set those line-solitons to meet at
the origin, and fix the amplitude of the soliton in $y<0$ to be 2.
$A_0$ is the amplitude of the soliton in $y>0$. The right panel
shows the incomplete chord diagrams corresponding to the choice of
the values $A_0$ and $\Psi_0$. There is no corresponding exact
soliton solution on the dotted line of $k_2=k_3$.} \label{fig:IV}
\end{figure}

The initial wave form is illustrated in the left panel of Figure
\ref{fig:IV}. The right figure shows the (incomplete) chord diagrams
corresponding to the initial V-shape waves; that is, the upper chord
represents the (semi-infinite) line-soliton in $y>0$ and the lower
one represents the (semi-infinite) line-soliton in $y<0$.  We have
done the numerical simulations with six different cases which are
marked  in Figure \ref{fig:IV} as  (a) through (f). The main result
in this section is to show that each {\it incomplete} chord diagram
representing the initial wave gets a completion with minimum
additional chords which represents the asymptotic pattern of the
solution. Here the {\it minimal completion} implies that for a given
incomplete chord diagram, one adds extra chords with minimum total
length so that the complete chord diagram represents a derangement
(see also \cite{KOT:09}). For our examples of the cases (a) through
(f), we have the following completions:
\begin{itemize}
\item[(a)]  With $k_1=k_2$, the completion is $\pi=\binom{1~3~4}{3~4~1}$, i.e. add $[3,4]$-chord in $y>0$.  This implies that $[3,4]$-soliton appears in the asymptotic solution in $y{{>}}0$, and
the corresponding exact solution is a $(1,2)$-soliton (cf. Figure \ref{fig:2}).
\item[(b)]  With $k_3=k_4$, the completion is $\pi=\binom{1~2~3}{3~1~2}$, i.e. add $[1,2]$-chord in $y<0$
which appears in the asymptotic solution, and the corresponding solution is a $(2,1)$-soliton (cf. Figure \ref{fig:2}).
\item[(c)] The completion is $\pi=(3142)$, i.e. add $[1,2]$-chord in $y<0$ and $[3,4]$-chord in $y>0$.  These corresponding solitons appear behind those initial solitons (cf. Figure \ref{fig:3142}).
\item[(d)] The completion is $\pi=(2143)$, i.e. add $[1,2]$-chord in $y<0$ and $[3,4]$-chord in $y>0$.
The soliton solution is of O-type, and the semi-infinite initial solitons extend to generate other half
with the phase shift (cf. Figure \ref{fig:2143}).
\item[(e)]  The {\it minimal} completion is $\pi=(2341)$, i.e. add $[1,2]$- and $[3,4]$-chords in $y>0$.
This is a $(1,3)$-soliton.  (Note that the initial wave is a front half of P-type soliton solution, but
the minimal completion is {\it not} of P-type.)
Since $[1,2]$-soliton in $y>0$ appears ahead of the wavefront consisting of $[2,3]$-soliton in $y>0$
and $[1,4]$-soliton in $y<0$, the generation of $[1,2]$-soliton is impossible.
The $[1,2]$-soliton is a {\it virtual} one, but it gives an influence on bending the $[1,4]$-soliton in $y>0$
 (see below for the details).
\item[(f)]  The {\it minimal} completion is $\pi=(4123)$, i.e. add $[1,2]$- and $[3,4]$-chords in $y<0$.
The is a $(3,1)$-type, but again we note that $[3,4]$-soliton in
$y<0$ is a virtual one (i.e. it locates ahead of the wavefront
similar to the previous case (e)). We also observe the bending of
$[1,4]$-soliton in $y>0$ under the influence of this virtual soliton
(see again below for the details).

\end{itemize}
%%%%%%%%%%%%%%%%%%%%%%%%%%%%%%%%%%%%%%%%%%
%%%%%%%%%%%%%%%%%%%%%%%%%%%%%%%%%%%%%%%%

\subsubsection{The case (a)}  This is a critical case with {{$k_1=k_2$.}}  The line-solitons of V-shape initial wave are $[1,3]$-soliton in $y>0$ and  $[1,4]$-soliton in $y<0$.  We take the following values for
the amplitudes and angles of the those solitons,
\[\left\{
\begin{array}{llll}
A_{[1,3]}=A_0=\frac{1}{2}, \quad & \tan\Psi_{[1,3]}=-\frac{1}{2}=-\tan\Psi_0, \\[1.0ex]
A_{[1,4]}=2,\quad &\tan\Psi_{[1,4]}=\frac{1}{2}=\tan \Psi_0.
\end{array}\right.
\]
%%%%%%%%%%%%%%%%%%%%%%%%%
With the formulas for the amplitude
$A_{[i,j]}=\frac{1}{2}(k_i-k_j)^2$ and the angle
$\tan\Psi_{[i,k]}=k_i+k_j$, we have the $k$-parameters $(k_1=k_2,
k_3,k_4)=(-\frac{3}{4},\frac{1}{4},\frac{5}{4})$. The corresponding
exact solution is of $(1,2)$-type.
\begin{figure}[t!]
\begin{centering}
\includegraphics[scale=0.63]{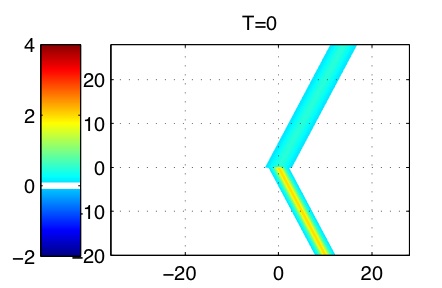}\includegraphics[scale=0.6]{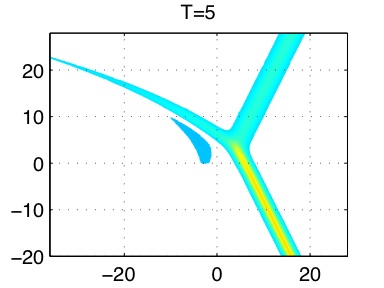}\includegraphics[scale=0.6]{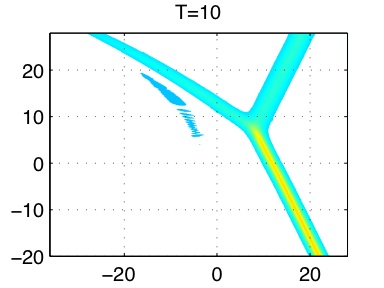}
\par\end{centering}
\begin{centering}
\includegraphics[scale=0.63]{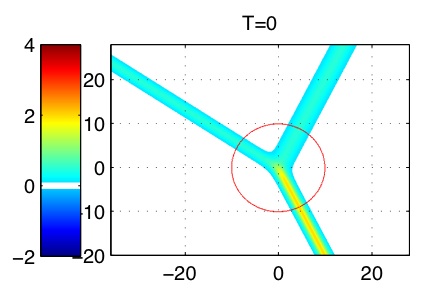}\includegraphics[scale=0.6]{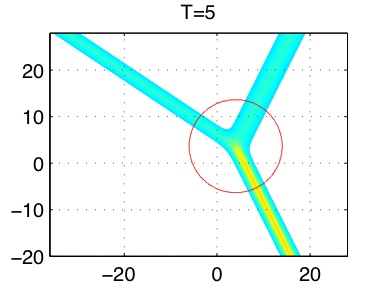}\includegraphics[scale=0.6]{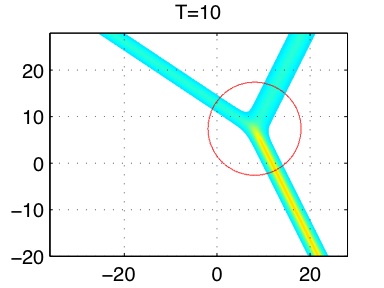}
\par\end{centering}
\begin{centering}
\includegraphics[scale=0.7]{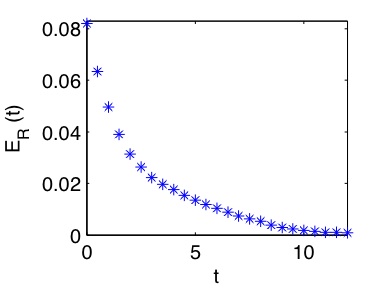}
\par\end{centering}
\caption{\label{fig:VE-a}Numerical simulation of the case (a) in Figure \ref{fig:IV}:   The initial wave consists of  $[1,3]$-soliton in $y>0$ and $[1,4]$-soliton in $y<0$. The top figures show the result of the direct simulation, the middle figures show the corresponding exact solution of $(1,2)$-type, and
the bottom one represents the error function $E(t)$ of (\ref{error}) which is minimized at $t=10$.   The circles in the exact solution shows the domain $D_r^t$.  A dispersive radiation appearing behind the interaction region may be considered as a reaction of the generation of the third soliton of $[3,4]$-type
as a result of resonant interaction (see the text for the details).}
\end{figure}
%%%%%%%%%%%%%%%%%%%%%%%%%%

The top figures of Figure \ref{fig:VE-a} illustrate the result of
the direct numerical simulation. The simulation shows a generation
of  third wave at the point of the intersection of those initial
solitons. This generation is  due to the resonant interaction, and
this may be identified as $[3,4]$-soliton of the exact solution,
that is, the resonance condition gives $[1,3]+[3,4]=[1,4]$ and
completes the incomplete chord diagram. The generation of
$[3,4]$-soliton may be also understood by decomposing the exact
solution into two parts, say $u_{\rm exact}=u+v$, where $u$ is the
solution of the KP equation with our initial wave of $[1,3]$- and
$[1,4]$-solitons. Then $v$ satisfies,
\[
\partial_x(4\partial_tv+6v\partial_xv+\partial_x^3v)+3\partial_y^2v = -6\partial_x^2(uv),
\]
which is the KP equation with one extra term in the right hand side.  If we assume that
the initial solitons remain as the part of
the exact solution $u_{\rm exact}$,  the product $uv$
in the extra term has nonzero value only near the origin at $t=0$
(this assumption is not totally correct, but the shifts of the initial solitons are very small
for most of the cases).
We then expect $v$ to be close to
the solution of the KP equation with the initial wave of $[3,4]$-soliton only for $y>0$ and $x<0$.
Since this $[3,4]$-soliton is missing in the front half $(x>0)$, the edge of the wave gets a strong
dispersive effect. Then the edge of the solution $v$ disperses away with a bow-shape wake
propagating toward the negative $x$-axis
as can be seen in the simulation (cf. Figure \ref{fig:VE-a}). Recall here that
the dispersive waves (i.e. non-soliton parts) propagate in the negative $x$-direction (see
Section \ref{sec:exact}).  Then the decay of the $v$ solution at the front edge
 implies the appearance of the $[3,4]$-soliton in the exact solution as one can see from
$u=u_{\rm exact}-v$.  This argument may be also applied to other cases as well.

Minimizing the error function $E(t)$ at $t=T_0=10$ with
(\ref{minimization}), we obtain the $A$-matrix of $(1,2)$-type,
\[
A=\begin{pmatrix}
1 &0 & -1.27 \\
0& 1 & 0.63
\end{pmatrix},
\]
The minimization is achieved by adjusting the locations of the $[1,3]$- and $[1,4]$-solitons
of the wavefront, and it gives the shifts
\[
x_{[1,3]}^+= -0.0079,   \qquad  x^+_{[1,4]}=-0.120.
\]
(The $A$-matrix is then obtained using (\ref{1-2locations})). Thus
the new line-soliton of $[3,4]$-type has a relatively large negative
shift. This may be explained as follows: The $[3,4]$-soliton is
generated as a part of the tail of $[1,4]$-soliton, and in the
beginning stage of the generation, the $[3,4]$-soliton has a small
amplitude and propagates with slow speed. This causes a negative
shift of this soliton. Also the negative shift in the
$[1,4]$-soliton is due to the generation of the $[3,4]$-soliton,
i.e. losing its momentum. The middle figures of Figure
\ref{fig:VE-a} show the exact solutions generated by the
$\tau$-function with those parameters $(k_1=k_2,
k_3,k_4)=(-\frac{3}{4},\frac{1}{4},\frac{5}{4})$ and the $A$-matrix
given above. The circles in the figures show $D_r^t$ where we take
$r=12$ and the center is given by
$(x_0(t)=\frac{13}{16}t-0.0637,~y_0(t)=\frac{3}{4}t-0.112)$. In the
bottom figures of Figure \ref{fig:VE-a}, we plot the error function
$E(t)$ of \eqref{error}.  Figure \ref{fig:VE-a} indicates that the
asymptotic solution seems to converge to the exact solution of
$(1,2)$-type with the above $A$-matrix and the same $k$-parameters.

\subsubsection{The case (b)}  This is also a critical case with $k_3=k_4$. The line-solitons of V-shape initial wave are $[1,3]$-soliton in $y>0$ and
$[2,3]$-soliton in $y<0$.  We take the amplitudes and angles of
those solitons to be
\[\left\{
\begin{array}{llll}
A_{[1,3]}=A_0=\frac{49}{8}, \quad & \tan\Psi_{[1,3]}=-\frac{3}{4}=-\tan\Psi_0, \\[1.0ex]
A_{[2,3]}=2,\quad &\tan\Psi_{[2,3]}=\frac{3}{4}=\tan \Psi_0.
\end{array}\right.
\]
This then gives the $k$-parameters $(k_1, k_2,
k_3=k_4)=(-\frac{17}{8},-\frac{5}{8},\frac{11}{8})$.
%%%%%%%%%%%%%%%%%%%%%%%%%%%%
\begin{figure}[t!]
\begin{centering}
\includegraphics[scale=0.63]{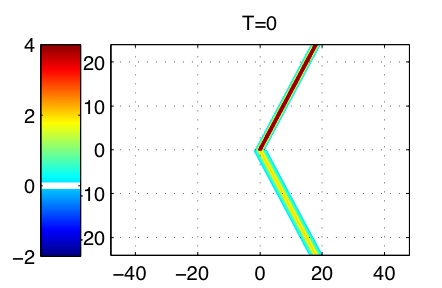}\includegraphics[scale=0.6]{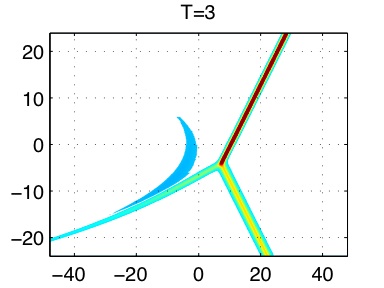}\includegraphics[scale=0.6]{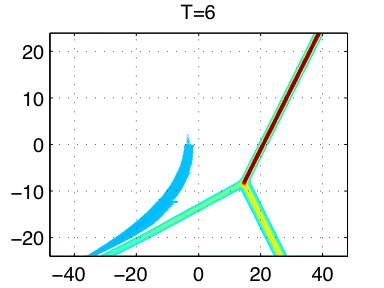}
\par\end{centering}
\begin{centering}
\includegraphics[scale=0.63]{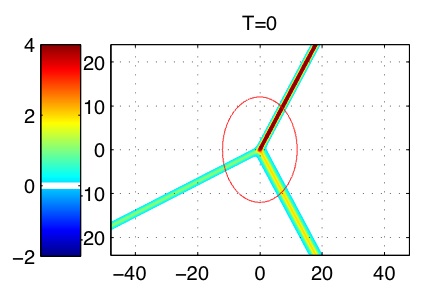}\includegraphics[scale=0.6]{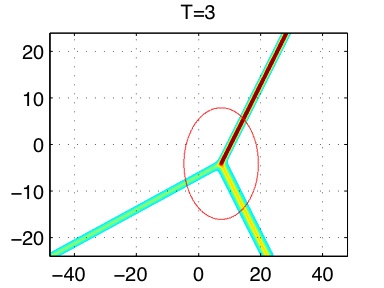}\includegraphics[scale=0.6]{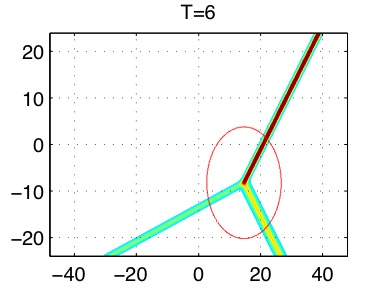}
\par\end{centering}
\begin{centering}
\includegraphics[scale=0.7]{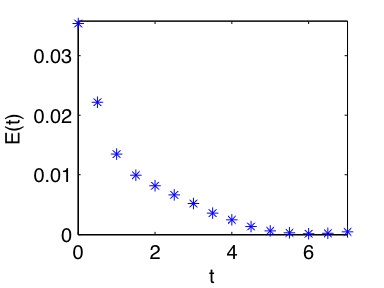}
\par\end{centering}

\caption{\label{fig:VE-b} Numerical simulation of the case (b) in Figure \ref{fig:IV}: The initial wave
consists of $[1,4]$- and $[2,4]$-solitons in $y>0$ and $y<0$ respectively.
The top figures show the result of the direct simulation, and the middle figures show the corresponding exact solution of $(2,1)$-type. The bottom graph shows the error function $E(t)$ of \eqref{error} which is minimized at $t=6$.}
\end{figure}
%%%%%%%%%%%%%%%%%%%%%%%%%%%%%%%%%%%%%%%
Figure \ref{fig:VE-b} illustrates the result of the numerical
simulation. The top figures show the direct simulation of the KP
equation. Notice again that a bow-shape wake behind the interaction
point corresponds to the $v$ solution in the decomposition,
$u=u_{\rm exact}-v$, and the decay of $v$ implies the appearance (or
generation) of $[1,2]$-soliton. Notice that $v$ gives the
$[1,2]$-soliton part in the exact solution. The middle figures show
the corresponding exact solution of $(2,1)$-type, whose $A$-matrix
is obtained by minimizing the error function $E(t)$ at $t=6$,
\[
A=\begin{pmatrix} 1& 1.15& 1.21\end{pmatrix}.
\]
The shifts of locations $x^+_{[i,j]}$ obtained in the minimization are given by
\[
x_{[1,3]}^+=-0.0545,\qquad x_{[2,3]}^+=-0.0254,
\]
using (\ref{2-1locations}). Those negative shifts may be explained
in the similar way as in the previous case. In those figures, the
domain $D_r^t$ has $r=12$ and its center
$(x_0(t)=\frac{157}{64}t-0.0399 , ~y_0(t)=-\frac{11}{8}t+0.0194 )$.
The bottom graph in Figure \ref{fig:VE-b} shows
 that the solution converges asymptotically to
the exact solution of $(2,1)$-type with the above $A$-matrix and the same $k$-parameters.

\subsubsection{The case (c)}  The line-solitons of the initial wave are $[1,3]$-soliton in $y>0$ and $[2,4]$-soliton in $y<0$.  We consider the case with the larger soliton in $y>0$, and take
the amplitudes and angles of those solitons as
\[\left\{
\begin{array}{llll}
A_{[1,3]}=A_0=2, \quad & \tan\Psi_{[1,3]}=-1=-\tan\Psi_0, \\[1.0ex]
A_{[2,4]}=2,\quad &\tan\Psi_{[2,4]}=1=\tan \Psi_0.
\end{array}\right.
\]
 The $k$-parameters are then given by $(k_1,k_2,k_3, k_4)=(-\frac{3}{2},-\frac{1}{2},\frac{1}{2},\frac{3}{2})$. The exact solution obtained by the completion of the chord diagram is given by
 $(3142)$-type soliton solution.
%%%%%%%%%%%%%%%%%%%%%%%%%%%%%%%%%%%%%%%
\begin{figure}[t!]
\begin{centering}
\includegraphics[scale=0.63]{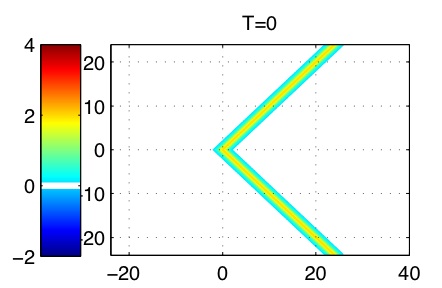}\includegraphics[scale=0.6]{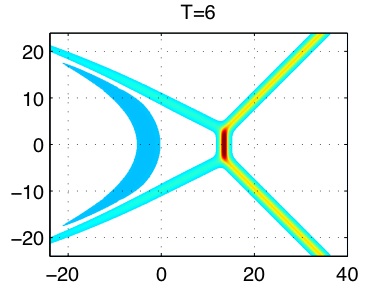}\includegraphics[scale=0.6]{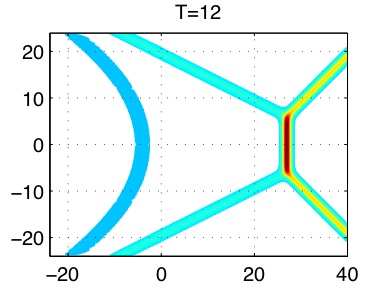}
\par\end{centering}
\begin{centering}
\includegraphics[scale=0.63]{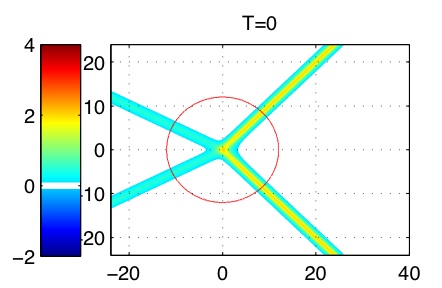}\includegraphics[scale=0.6]{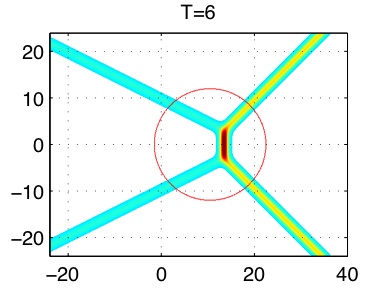}\includegraphics[scale=0.6]{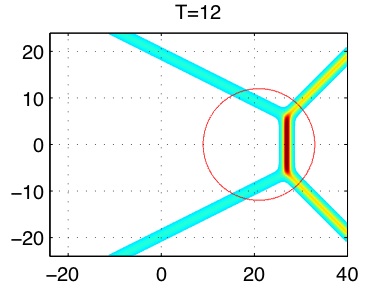}
\par\end{centering}
\begin{centering}
\includegraphics[scale=0.7]{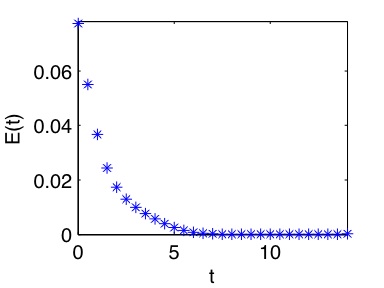}
\par\end{centering}\caption{\label{fig:VE-e}  Numerical simulation of the case (c) in Figure \ref{fig:IV}: The initial wave consists of  $[1,3]$-soliton in $y>0$ and $[2,4]$-soliton in $y<0$. The top figures show the result of the direct simulation, and the middle figures show the corresponding exact solution of $({{3142}})$-type.
The bottom graph of the error function $E(t)$ which is minimized at $t=10$.
A large amplitude intermediate soliton is generated at the intersection point, and it corresponds
to the $[1,4]$-soliton with the amplitude $A_{[1,4]}=4.5$. }
\end{figure}
%%%%%%%%%%%%%%%%%%%%%%%%%%%%%%%%%%%%%%%%%
Figure \ref{fig:VE-e} illustrates the result of the numerical
simulation. The top figures show the direct simulation of the KP
equation. We observe a bow-shape wake behind the interaction point.
The wake expands and decays, and then we see the appearance of new
solitons which form resonant interactions with the initial solitons.
One should note that the solution generates a large amplitude
intermediate soliton at the interaction point, and this soliton is
identified as $[1,4]$-soliton with the amplitude $A_{[1,4]}=4.5.$
This solution has been considered to describe the Mach reflection of
shallow water waves, and the $[1,4]$-soliton describes the wave
called the Mach stem \cite{CK:09, KOT:09}.

The middle figures in Figure \ref{fig:VE-e} show the corresponding
exact solution of $(3142)$-type whose $A$-matrix is found by
minimizing $E(t)$ at $t=10$,
\[
A=\begin{pmatrix}
1 & 1.92  &0&  -1.96\\
0& 0 &1 & 0.64
\end{pmatrix}
\]
In the minimization to find the $A$-matrix, we adjust the locations
of solitons including the intermediate solitons, and using
(\ref{3142A}), we obtain
\[
 x_{[1,3]}^+=x_{[2,4]}^+=-0.01,\qquad s=0.980.
 \]
Those values indicate that the solution is very close to the exact solution for all the time.
The negative value of the shifts $x^+_{[i,j]}$ is due to the generation of  a large amplitude
soliton $[1,4]$-type (i.e. the initial solitons loose their momentum), and $s<1$ implies that the $[1,4]$-soliton is  generated after $t=0$.
Also note that the
$[1,4]$-soliton now resonantly interact with $[1,3]$- and
$[2,4]$-solitons to create new solitons $[1,2]$- and
$[3,4]$-solitons (called the reflected waves as considered in the
Mach reflection problem \cite{M:77, CK:09, KOT:09}).  This process
then seems to compensate the shifts of incident waves, even though
we observe a large wake behind the interaction point. Notice that
the wake disperses, and the exact solution starts to appear as the
wake decays (recall again the decomposition $u_{\rm exact}=u+v$
(where $v$ is the wake and $u$ is the solution).

The bottom graph in Figure \ref{fig:VE-e} shows a rapid convergence
of the initial wave to $(3142)$-type soliton solution with those
parameters of the $A$-matrix and $k$ values given above, and for the
error function, we use $D^t_r$ with $r=12$.

\subsubsection{The case (d)}  The line-solitons of the initial wave are $[1,2]$-soliton in $y>0$ and  $[3,4]$-soliton in $y<0$.  This case corresponds to the solitons with a large angle interaction.
For simplicity, we consider a symmetric initial data with the
initial wave whose amplitudes and angles are given by
\[\left\{
\begin{array}{llll}
A_{[1,2]}=A_0=2, \quad & \tan\Psi_{[1,2]}=-\frac{12}{5}=-\tan\Psi_0, \\[1.0ex]
A_{[3,4]}=2,\quad &\tan\Psi_{[3,4]}=\frac{12}{5}=\tan \Psi_0.
\end{array}\right.
\]
The $k$-parameters are then given by $(k_1, k_2,k_3 , k_4)=(-\frac{11}{5},
-\frac{1}{5}, \frac{1}{5}, \frac{11}{5})$. The completion of the chord diagram gives O-type
soliton solution, i.e. $(2143)$-type. This implies that the solution
extends the initial solitons to the negative $x$-direction with the
same types, i.e. $[1,2]$- and $[3,4]$-types in $y<0$ and $y>0$,
respectively. One should note that the solution also generates the
phase shift determined uniquely by the $k$-parameters.
%%%%%%%%%%%%%%%%%%%%%%%%%%%%%%%%%%%%%%%%%%%%
\begin{figure}[t!]
\begin{centering}
\includegraphics[scale=0.63]{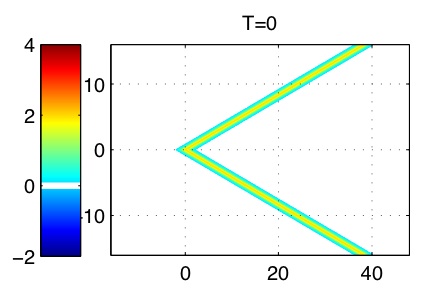}\includegraphics[scale=0.6]{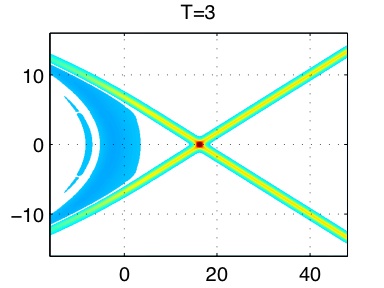}\includegraphics[scale=0.6]{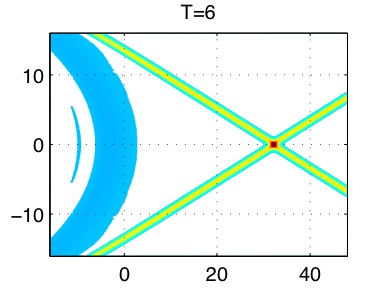}
\par\end{centering}
\begin{centering}
\includegraphics[scale=0.63]{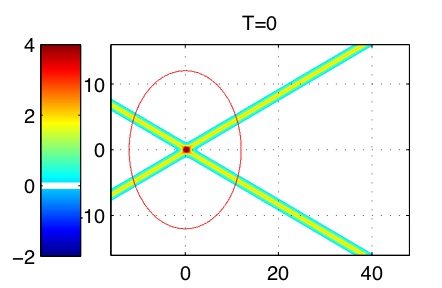}\includegraphics[scale=0.6]{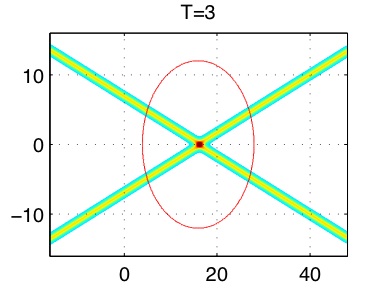}\includegraphics[scale=0.6]{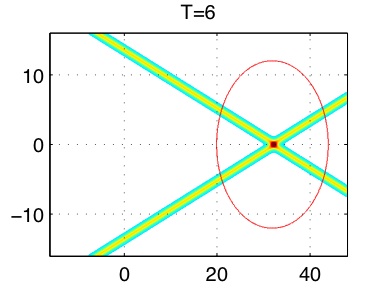}
\par\end{centering}
\begin{centering}
\includegraphics[scale=0.7]{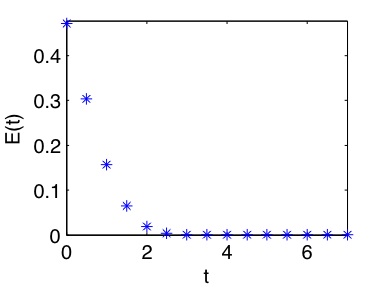}
\par\end{centering}
\caption{\label{fig:VE-f} Numerical simulation of the case (d) in
Figure \ref{fig:IV}. The initial wave consists of $[1,2]$-soliton in
$y>0$ and $[3,4]$-soliton in $y<0$. The upper figures show the
result of the direct simulation, and the middle figures show the
corresponding exact solution of $(2143)$-type (O-type). Notice a
large wake behind the interaction point which extends the initial
solitons. The bottom graph shows the error function $E(t)$ which is
minimized at $t=6$.}
\end{figure}
%%%%%%%%%%%%%%%%%%%%%%%%%%%%%%%%%%%%%%%%%%%%%%%%
Figure \ref{fig:VE-f} illustrates the result of the numerical
simulation. The top figures show the direct simulation of the KP
equation. The wake behind the interaction point has a large negative
amplitude, and this corresponds to the $v$ solution in $u_{\rm
exact}=u+v$. The middle figures show the corresponding O-type exact
solution whose $A$-matrix is determined by minimizing the error
function $E(t)$ at $t=10$,
\[
A=\begin{pmatrix}
1 & 1.91 & 0 & 0\\
0 & 0 & 1 &  0.17
\end{pmatrix}
\]
Because of our symmetric initial data, we consider the shifts
$x^+_{[1,2]}=x^+_{[3,4]}$ for the minimization, and using
(\ref{OA}), we obtain
\[
x_{[1,2]}^+=x_{[3,4]}^+=-0.020.
\]
The negative shifts imply the slow-down of the incidence waves due
to the generation of the solitons extending the initial solitons in
the negative $x$-direction.  Those new parts of solitons have the
locations $x^-_{[i,j]}$. The phase shifts $\Delta
x_{[i,j]}=x^-_{[i,j]}-x^+_{[i,j]}$ for the line-solitons are
calculated from (\ref{Oshift}), and they are
\[
\Delta x_{[1,2]}=\Delta x_{[3,4]}=0.593.
\]
The positivity of the phase shifts is due to the attractive force
between the line-solitons, and this explains the slow-down of the
initial solitons, i.e. the small negative shifts of $x_{[i,j]}^+$.
The bottom graph in Figure \ref{fig:VE-f} shows $E(t)$ which is
minimized at $t=6$, and we take $r=12$ for the domain $D_r^t$. One
can see a rapid convergence of the solution to the O-type exact
solution with those parameters. One should however remark that for a
case near the critical one with $k_2=k_3$, there exists a large
phase shift in the soliton solution, and  the convergence is very
slow. Note that the case with $k_2=k_3$ also corresponds to the
boundary of the previous case (c) (See Figure \ref{fig:IV}), and the
amplitude of the intermediate soliton generated at the intersection
point reaches {\it four} times larger than the initial solitons at
this limit. This large amplitude wave generation has been considered
as the Mach reflection problem of shallow water wave \cite{M:77,
PTLO:05, CK:09, KOT:09, TOM:08}.

\subsubsection{The case (e)}  The line-solitons of the initial wave are $[2,3]$-soliton in $y>0$ and
$[1,4]$-soliton in $y<0$.  We consider the solitons parallel to the $y$-axis with
the amplitudes,
\[\left\{
\begin{array}{llll}
A_{[2,3]}=A_0=0.25, \quad & \tan\Psi_{[2,3]}=0=-\tan\Psi_0, \\[1.0ex]
A_{[1,4]}=2,\quad &\tan\Psi_{[1,4]}=0=\tan \Psi_0.
\end{array}\right.
\]
Then the $k$-parameters are given by $(k_1, k_2, k_3,k_4)=(-1,-\frac{1}{2\sqrt{2}},\frac{1}{2\sqrt{2}},1)$.
We also considered the case with nonzero angle (i.e. oblique to the $y$-axis with $\Psi_0>0$), and
we expected to see a P-type soliton as the asymptotic solution.  However, the
numerical simulation suggests that we do not see a solution of P-type, but the asymptotic solution is somewhat close to the exact solution of $(2341)$-type as obtained by the minimal completion of the chord diagram.
The $(2341)$-type is a $(1,3)$-type soliton solution with $[1,4]$-soliton in $y<0$ and
$[1,2]$-, $[2,3]$- and $[3,4]$-solitons in $y>0$.  Here one should note that the $[1,2]$-soliton in $y>0$
locates ahead of the wavefront consisting of $[1,4]$- and $[2,3]$-solitons, so that this soliton cannot be
generated by the resonant interaction of those $[1,4]$- and $[2,4]$-solitons.

\begin{figure}
\begin{centering}
\includegraphics[scale=0.63]{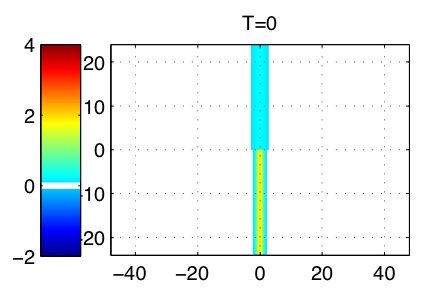}\includegraphics[scale=0.6]{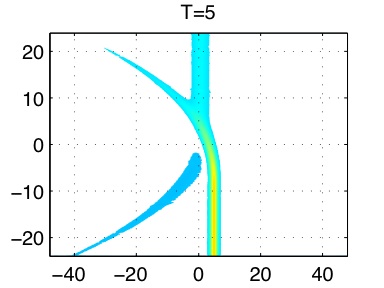}\includegraphics[scale=0.6]{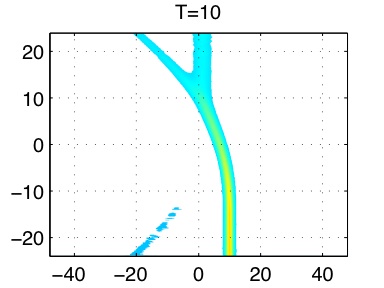}
\par\end{centering}
\begin{centering}
\includegraphics[scale=0.63]{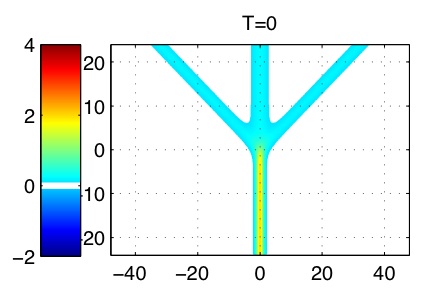}\includegraphics[scale=0.6]{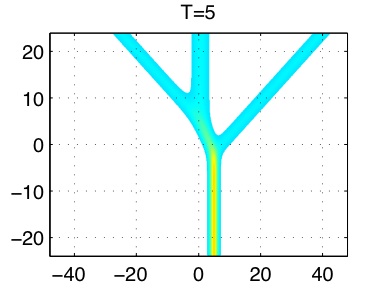}\includegraphics[scale=0.6]{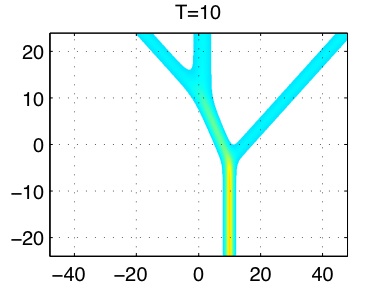}
\par\end{centering}
%\begin{centering}
%\includegraphics[scale=0.5]{VU4_0_0_err_}
%\par\end{centering}
\caption{\label{fig:VE-c}Numerical simulation of the case (e) in Figure \ref{fig:IV}: The initial wave consists of  $[2,3]$-soliton in $y>0$ and $[1,4]$-soliton in $y<0$. The upper figures show the result of the direct simulation, and the lower figures show the corresponding exact solution of $(2341)$-type obtained by the {\it minimal }completion
of the chord diagram given by the initial wave. Notice that the bending on the initial $[1,4]$-soliton
appears near the point where the {\it virtual} $[1,2]$-soliton intersects.
The dispersive radiation behind the intersection point may be the part of this $[1,4]$-soliton.}
\end{figure}

Figure \ref{fig:VE-c} illustrates the result of the numerical
simulation. The top figures show the direct simulation of the KP
equation. We observe that the $[1,4]$-soliton in $y<0$ propagates
faster and its tail extends to $y>0$. Also note that a large
dispersive wake appearing in $y<0$ has negative amplitude and
disperses in the negative $x$-direction. On the other hand, the tail
part in $y>0$ seems to have a resonant interaction with the
$[2,3]$-initial soliton and becomes a new soliton.  This behavior of
the solution may be explained by the chord diagram of $(2341)$-type
whose corresponding exact solution is illustrated in the bottom
figures in Figure \ref{fig:VE-c}:  The new soliton generated in
$y>0$ may be identified as $[3,4]$-soliton. Also we note that the
bending of the $[1,4]$-soliton observed in the simulation seems to
appear at the point where the $[1,2]$-soliton has a resonant
interaction with $[1,4]$-soliton. This means that the intermediate
soliton in the exact solution is $[2,4]$-soliton, and this soliton
may give a good approximation of the bending part of the
$[1,4]$-soliton.  It is interesting to consider this $[1,2]$-soliton
as a {\it virtual} one, which is not visible but has an influence to
the initial solitons. This can be also understood by the similar
argument using the decomposition $u_{\rm exact}=u+v$. In this case,
the initial data for $v$ is given by the $[1,2]$- and
$[3,4]$-solitons in $u_{\rm exact}$. Then the $[3,4]$-soliton in
$u_{\rm exact}$ appears when the solution $v$ decays in this part.
On the other hand, the decay of $v$ in the $[1,2]$-soliton part
seems to occur only in the region behind the front solitons as seen
at $t=5$. Non-decaying of the $[1,2]$-soliton implies that
$[1,2]$-soliton never appears in the solution $u$, because of the
cancellation of $u_{\rm exact}$ and $v$ on this soliton part. We
will give more details in a future communication.
%%%%%%%%%%%%%%%%%%%%%%%%%%%%%%%%%%%%%%%%%%%%
%%%%%%%%%%%%%%%%%%%%%%%%%%%%%%%%%%%%%%%%%%%%%%%%

\subsubsection{The case (f)}  The line-solitons of the initial wave are $[1,4]$-soliton in $y>0$ and $[2,3]$-soliton in $y<0$. Contrary to the previous case, we have now the larger soliton in $y>0$
and the smaller one in $y<0$.  In a comparison, we also consider the solitons parallel to the $y$-axis. We then take the amplitudes and angles of the those solitons as
\[\left\{
\begin{array}{llll}
A_{[1,4]}=A_0=6, \quad & \tan\Psi_{[1,4]}=0=-\tan\Psi_0, \\[1.0ex]
A_{[2,3]}=2,\quad &\tan\Psi_{[2,3]}=0=\tan \Psi_0.
\end{array}\right.
\]
Then the $k$-parameters are given by $(k_1, k_2, k_3, k_4)=(-\sqrt{3},-1,1,\sqrt{3})$.
Since the upper $[1,4]$-soliton propagates faster than the lower one, the lower tail of this soliton
appears and then resonates with the lower $[2,3]$-soliton.
Following the similar arguments as in the previous case,
we expect  that $(3,1)$-soliton is the corresponding exact solution which is of $(4123)$-type as
given by the minimal completion of the chord diagram.
%%%%%%%%%%%%%%%%%%%%%%%%%%%%%%%%%%%
\begin{figure}[t!]
\begin{centering}
\includegraphics[scale=0.63]{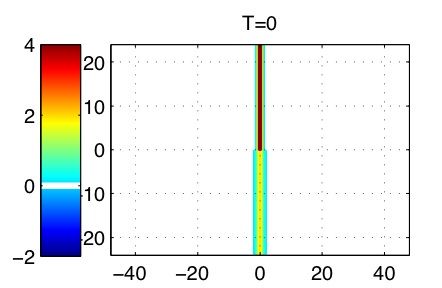}\includegraphics[scale=0.6]{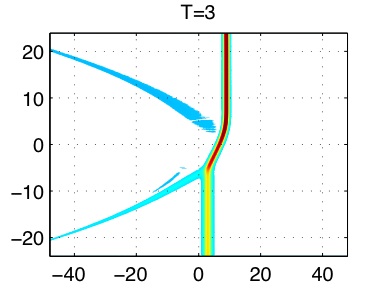}\includegraphics[scale=0.6]{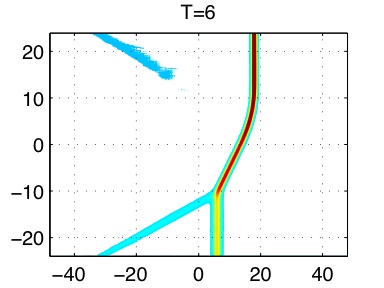}
\par\end{centering}
\begin{centering}
\includegraphics[scale=0.63]{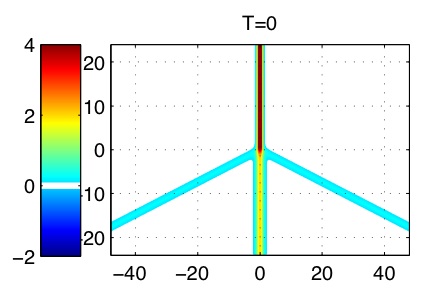}\includegraphics[scale=0.6]{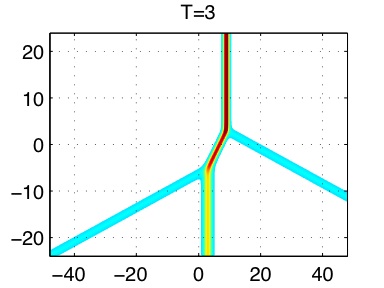}\includegraphics[scale=0.6]{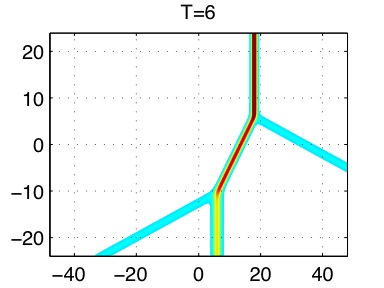}
\par\end{centering}
%\begin{centering}
%\includegraphics[scale=0.5]{VU6_0_0_err_}
%\par\end{centering}
\caption{\label{fig:VE-d} Numerical simulation of the case (f) in Figure \ref{fig:IV}: The initial wave consists of  $[1,4]$-soliton in $y>0$ and $[2,3]$-soliton in $y<0$. The upper figures show the result of the direct simulation, and the lower figures show the corresponding exact solution of $(4123)$-type obtained by
the minimal completion of the chord diagram. Notice that the bending in the $[1,4]$-soliton appears near
the intersection point with the {\it virtual} $[3,4]$-soliton, and the dispersive radiation observed in the simulation
is a part of this $[3,4]$-soliton.}
\end{figure}
%%%%%%%%%%%%%%%%%%%%%%%%%%%%%%%%%%%%%%%%%%%%%%
Figure \ref{fig:VE-d} illustrates the result of the numerical
simulation. The upper figures show the direct simulation, and the
lower figures show the exact solution of $(4123)$-type. As in the
previous case (e), this case shows that the $[3,4]$-soliton part in
the exact solution does not appear in the solution of the
simulation, that is, the $[3,4]$-soliton is a virtual one. This can
be also explained with the similar argument as in the previous case.
Notice again that the $[1,4]$-soliton in $y>0$ has a virtual
interaction with $[3,4]$-soliton to get bending near $(x=18,y=7)$ at
$t=6$.

\subsection{$X$-shape initial waves}

\begin{figure}[t!]
\centering
\includegraphics[scale=0.54]{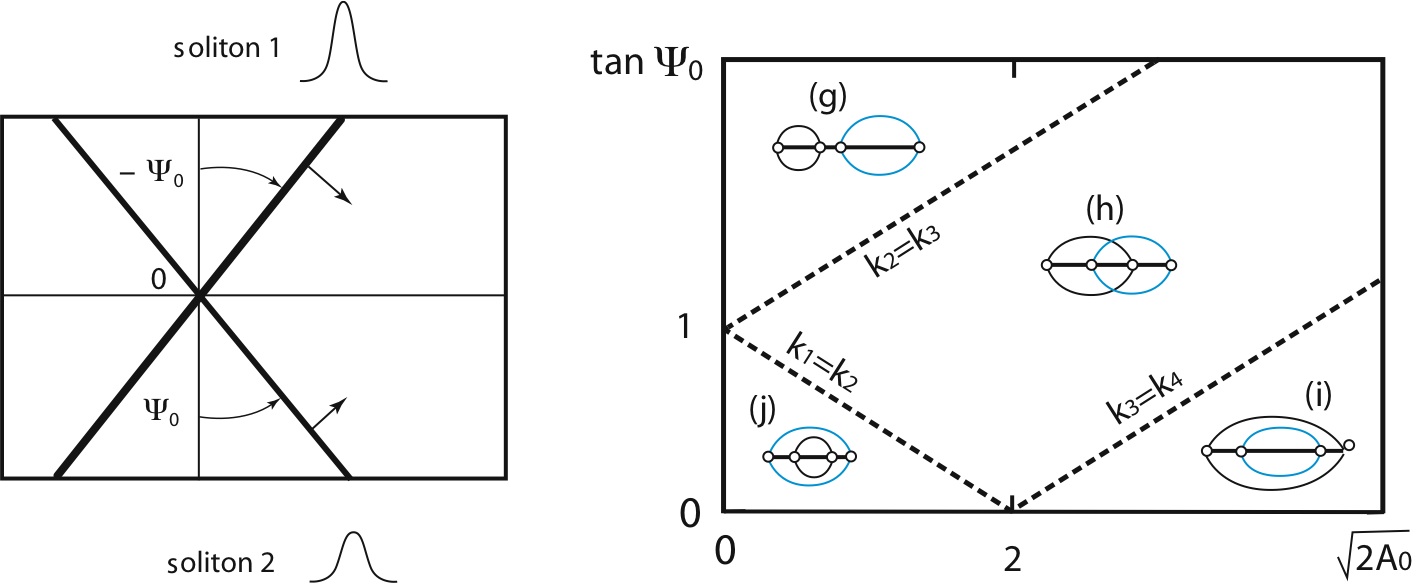}
\caption{X-shape initial waves. Each line of the X-shape is an
infinite line-soliton solution. We set those line-solitons to meet
at the origin, and fix the amplitude of the soliton with positive
angle $\Psi_0$ to be 2 (the corresponding chord is shown in the
lighter color in each region). $A_0$ is the amplitude of the soliton
with negative angle $-\Psi_0$. The right panel shows the
corresponding chord diagrams with the choice of the values $A_0$ and
$\Psi_0$. There is no exact solution for the degenerate cases marked
by the dotted lines.} \label{fig:X}
\end{figure}

The initial wave form is illustrated in the left panel of Figure
\ref{fig:X}.  The right figure shows the chord diagrams
corresponding to the initial waves with the parameters
$(A_0,\tan\Psi_0)$. We carried out four different cases marked with
(g) through (j) in Figure \ref{fig:X}.  The main result in this
section is to show that each simple sum of line-solitons gets
asymptotically to the exact solution corresponding to the chord
diagram given by the initial wave. One should however note that
those initial waves are not the exact solutions, and the simulations
provide stability analysis for those exact solutions.  For example,
if the exact solution has the phase shift, the asymptotic solution
actually converges to this exact solution by generating the correct
phase shifts.
%%%%%%%%%%%%%%%%%%%%%%%%%%%%%%%%%%%%%%%%%%
%%%%%%%%%%%%%%%%%%%%%%%%%%%%%%%%%%%%%%%%

\subsubsection{The case (g)}
\begin{figure}[t!]
\begin{centering}
\includegraphics[scale=0.63]{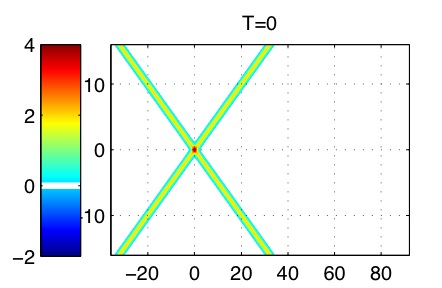}\includegraphics[scale=0.6]{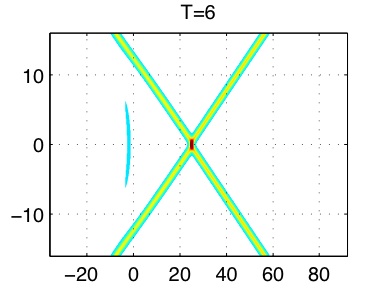}\includegraphics[scale=0.6]{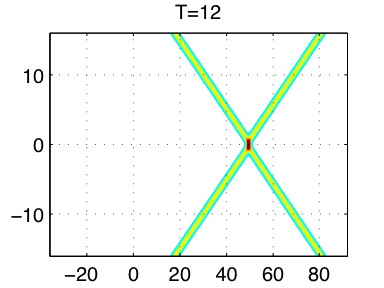}
\par\end{centering}
\begin{centering}
\includegraphics[scale=0.63]{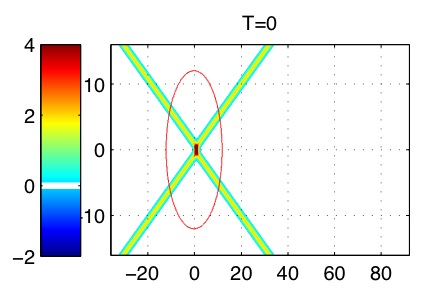}\includegraphics[scale=0.6]{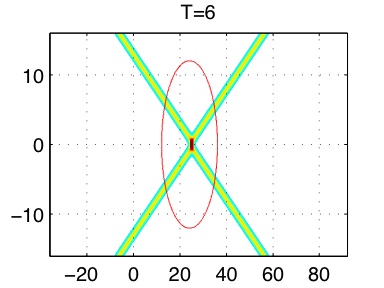}\includegraphics[scale=0.6]{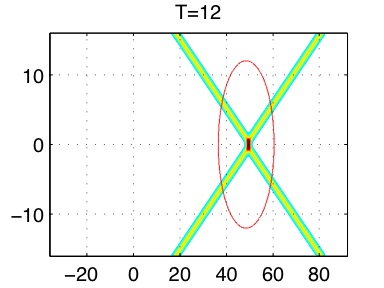}
\par\end{centering}
\begin{centering}
\includegraphics[scale=0.7]{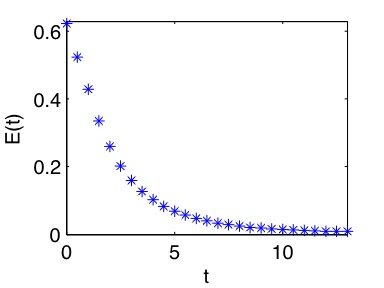}
\par\end{centering}

\caption{\label{fig:X-g} Numerical simulation of the case (g) in Figure \ref{fig:X}.
The initial wave is the sum of $[1,2]$- and $[3,4]$-solitons.  The top figures show the numerical
simulation, and the middle figures show the corresponding exact solution of $(2143)$-type, i.e. O-type.
The bottom graph shows $E(t)$ of \eqref{error} which is minimized at $t=12$.
Notice that a small dispersive radiation appears to generate the correct phase shifts.}
\end{figure}

The initial wave consists of $[1,2]$-  and $[3,4]$-soliton.  To
emphasize the phase shift appearing in the exact solution of
$(2143)$-type, we consider the initial solitons to be close to the
boundary at $k_2=k_3$ (but we cannot take the values so close, since
the generation of a large phase-shift needs a long time
computation).  For simplicity, we also take a symmetric initial wave
whose
 amplitudes and angles of the initial solitons are given by
\[\left\{
\begin{array}{llll}
A_{[1,2]}=A_0=2,  &  \tan\Psi_{[1,2]}=-2.02=-\tan\Psi_0\\[1.0ex]
A_{[3,4]}=2 & \tan\Psi_{[3,4]}=2.02=\tan\Psi_0
\end{array}
\right.
\]
The $k$-parameters are then given by
$(k_1,k_2,k_3,k_4)=(-2.01,-0.01,0.01,2.01)$.
 The numerical simulation in Figure \ref{fig:X-g} demonstrates
the generation of the phase shifts after the interaction,  that is, the line-solitons in the left side
are accelerated by their interaction. Note here that the interaction generates the dispersive wake behind
the interaction point which
propagates in the negative $x$-direction.

Minimizing the error function $E(t)$ at $t=12$, we obtain the
$A$-matrix,
\[
A=\begin{pmatrix}
1 &  3.029  &  0  &  0 \\
0 &   0  & 1  &  0.015
\end{pmatrix}
\]
In the minimization by adjusting the locations of the solitons, we use $x^+_{[1,2]}=x^+_{[3,4]}$
(due to the symmetry of the solution), and then we obtain
\[x_{[1,2]}^+=x_{[3,4]}^+=-0.21.
\]
The negative values are due to the generation of the positive phase shifts given by (\ref{Oshift}),
\[
\Delta x_{[1,2]}=\Delta x_{[3,4]}=1.963 >0.
\]
Namely the accelerations of the back parts of the line-solitons
(i.e. the positive phase shifts) imply the deceleration of the front
parts of solitons. The middle figures in Figure \ref{fig:X-g} shows
the corresponding O-type soliton solution with this $A$-matrix. The
circle shows the domain $D_r^t$ with $r=12$. Notice the dispersive
wake behind the interaction point in the simulation. The bottom
graph in Figure \ref{fig:X-g} shows the convergence of the solution
to the O-type solution given by those $A$-matrix and the same
$k$-parameters.

%%%%%%%%%%%%%%%%%%%%%%%%%%%%%%%%%%%
%%%%%%%%%%%%%%%%%%%%%%%%%%%%%%%%%%%%%%%

\subsubsection{The case (h)}
The initial wave is the sum of $[1,3]$-  and $[2,4]$-soliton. The
corresponding chord diagram is of $(3412)$-type (i.e. T-type), and
we expect the generation of a box at the intersection point of those
solitons. For simplicity, we consider a symmetric initial wave, and
take the amplitudes and angles of those solitons to be
\[\left\{
\begin{array}{llll}
A_{[1,3]}=A_0=2,  &  \tan\Psi_{[1,3]}=-1=-\tan\Psi_0\\[1.0ex]
A_{[2,4]}=2 & \tan\Psi_{[2,4]}=1=\tan\Psi_0
\end{array}
\right.
\]
 The $k$-parameters are then given by
$(k_1,k_2,k_3,k_4)=(-\frac{3}{2}, -\frac{1}{2},\frac{1}{2},\frac{3}{2})$.
Although T-type soliton appears for smaller angle $\Psi_0$, one should not take so small value.
For an example of the symmetric case, if we take $\Psi_0=0$, we obtain the KdV 2-soliton solution
with different amplitudes.
So for the case with very small angle $\Psi_0$, we expect to see those KdV solitons near the intersection point.
However, the solitons expected from the chord diagram have almost the same amplitude
as the incidence solitons for the case with a small angle.
The detailed study also shows that near the intersection point for T-type solution at the time when
all four solitons meet at this point (i.e X-shape), the solution has a small amplitude due to the
repulsive force similar to the KdV solitons (see also \cite{C:09}, where the initial X-shape wave with small angle
generates a large soliton at the intersection point). This then implies that our initial wave given by
the sum of two line-solitons creates a large
dispersive perturbation at the intersection point (this can be also seen in the next cases where we discuss the
P-type solutions).

\begin{figure}[t!]
\begin{centering}
\includegraphics[scale=0.63]{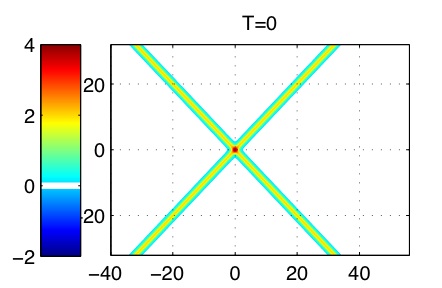}\includegraphics[scale=0.6]{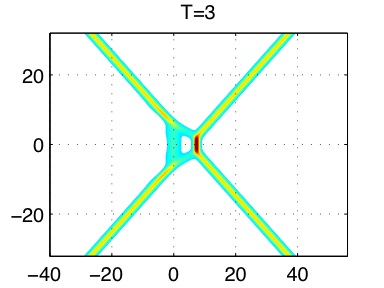}\includegraphics[scale=0.6]{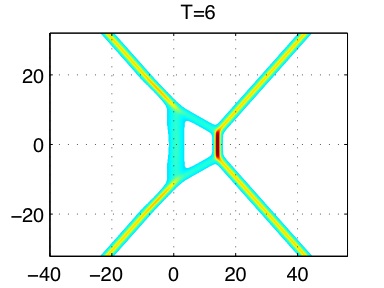}
\par\end{centering}
\begin{centering}
\includegraphics[scale=0.63]{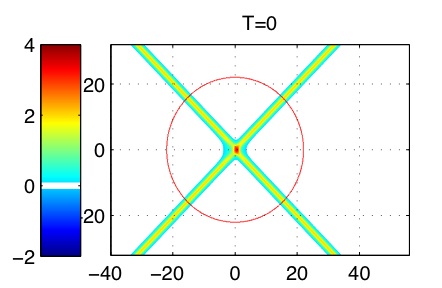}\includegraphics[scale=0.6]{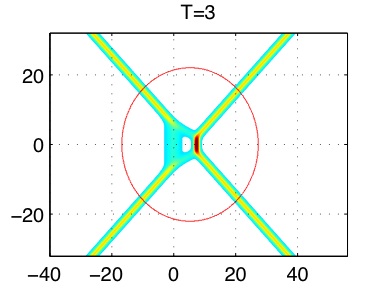}\includegraphics[scale=0.6]{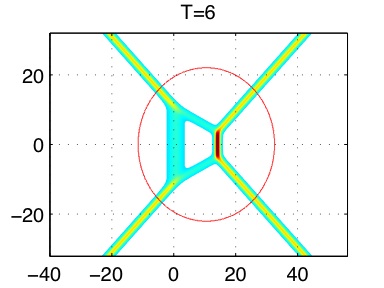}
\par\end{centering}
\begin{centering}
\includegraphics[scale=0.7]{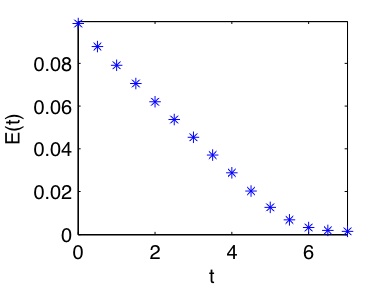}
\par\end{centering}\caption{\label{fig:X-h} Numerical simulation of the case (h) in Figure \ref{fig:X}.
The initial wave is the sum of $[1,3]$- and $[2,4]$-solitons.  The top figures show the numerical
simulation, and the middle figures show the corresponding exact solution of $(3412)$-type, i.e. T-type.
The bottom graph shows the error function $E(t)$ of \eqref{error} which is minimized at $t=6$.}
\end{figure}

The top figures in Figure \ref{fig:X-h} illustrate the numerical
simulation, which clearly shows an opening of a resonant box as
expected by the chord diagram of T-type. The corresponding exact
solution is illustrated in the middle figures, where the $A$-matrix
of the solution is obtained by minimizing the error function $E(t)$
of \eqref{error} at $t=6$,
 \[
 A=\begin{pmatrix}
 1 & 0 &  -0.368  & -0.330 \\
 0 & 1 & 1.198  & 0.123
 \end{pmatrix}
 \]
In the minimization, we take $x_{[1,3]}^{\pm}=x_{[2,4]}^{\pm}$ due to the symmetric profile of the solution,
and adjust the on-set of the box (recall (\ref{TA}), and note that the symmetry reduces the number of
free parameters to three).  We obtain
\[
x_{[1,3]}^+= x_{[2,4]}^+=0.025, \qquad r=3.63, \qquad s=0.350.
\]
The positive shifts of those $[1,3]$- and $[2,4]$-solitons in the wavefront indicate also the positive shift
of the newly generated soliton of $[1,4]$-type at the front. This is due to the repulsive force exists in
the KdV type interaction explained above, that is,
the interaction part in the initial wave has a larger amplitude than that of the exact solution, so that
this part of the solution moves faster than that in the exact solution. This difference may result as a shift
of the location
of the $[1,4]$-soliton. The phase shifts of the initial waves are obtained from (\ref{Tshift}),
\[
\Delta x_{[1,3]}=\Delta x_{[2,4]}=0.549.
\]
The relatively large value $r>1$ indicates that the on-set of the box is actually much earlier that $t=0$,
and $s<1$ shows the positive phase shifts as calculated above.

The bottom graph in Figure \ref{fig:X-h} shows the evolution of the
error function $E(t)$ of \eqref{error} which is minimized at $t=6$.
Note here that the circular domain $D_r^t$ with $r=22$ covers well
the main feature of the interaction patterns for all the time
computed for $t\le 7$.

 %%%%%%%%%%%%%%%%%%%%%%%%%%%%%%%%%%%%%%%%%%%%%%
%%%%%%%%%%%%%%%%%%%%%%%%%%%%%%%%%%%%%%%%%%%%%%%

\subsubsection{The case (i)}
The initial wave consists of $[1,4]$-  and $[2,3]$-soliton.  The corresponding diagram is of $(4321)$-type (P-type), and the exact solution of this type has negative (repulsive) phase shift.
Here we show how the phase shift is generated in the simulation.
We take the amplitudes and angles
of those solitons to be
\[\left\{
\begin{array}{llll}
A_{[2,3]}=A_0=\frac{8}{25},  &  \tan\Psi_{[2,3]}=-0.4=-\tan\Psi_0\\[1.0ex]
A_{[1,4]}=2 & \tan\Psi_{[1,4]}=0.4=\tan\Psi_0
\end{array}
\right.
\]
The $k$-parameters are then given by
$(k_1,k_2,k_3,k_4)=(-\frac{4}{5},-\frac{3}{5},\frac{1}{5},\frac{6}{5})$.
%%%%%%%%%%%%%%%%%%%%%%%%%%%%%%%%%%%%%%%%%%%%%
\begin{figure}[t!]
\begin{centering}
\includegraphics[scale=0.63]{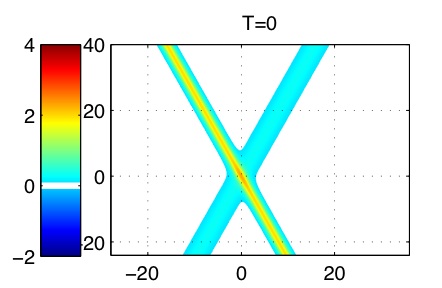}\includegraphics[scale=0.6]{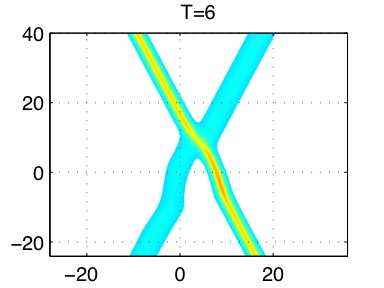}\includegraphics[scale=0.6]{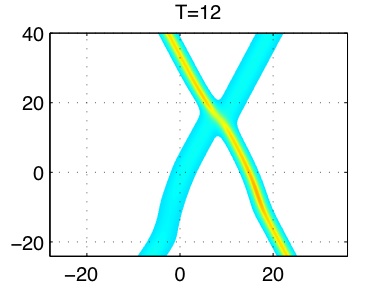}
\par\end{centering}
\begin{centering}
\includegraphics[scale=0.63]{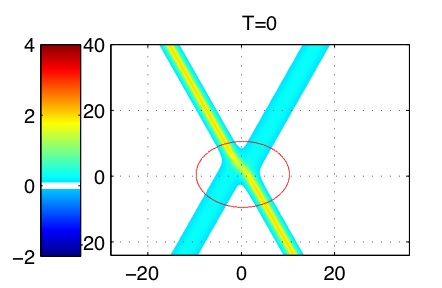}\includegraphics[scale=0.6]{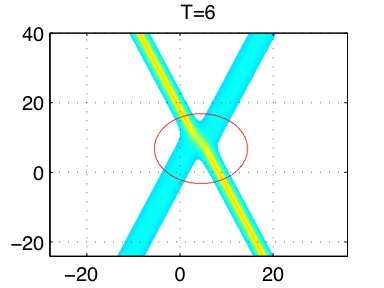}\includegraphics[scale=0.6]{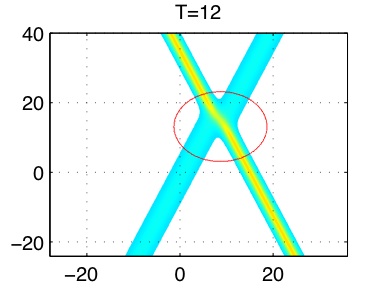}
\par\end{centering}
\begin{centering}
\includegraphics[scale=0.7]{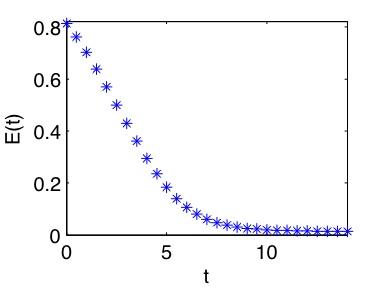}
\par\end{centering}
\caption{\label{fig:X-i} Numerical simulation of the case (i) in
Figure \ref{fig:X}. The initial wave is the sum of $[1,4]$- and
$[2,3]$-solitons.  The top figures show the numerical simulation,
and the middle figures show the corresponding exact solution of
$(4321)$-type, i.e. P-type. The bottom graph shows the error
function $E(t)$ of \eqref{error} which is minimized at $t=12$. Note
that the bending on the solitons disperses and the interaction
pattern converges locally to that of the corresponding exact
solution.}
\end{figure}
%%%%%%%%%%%%%%%%%%%%%%%%%%%%%%%%%%%%%%%%%%%%%%
In this case, we also note that taking a very small angle $\Psi_0$
may give a large perturbation to the exact solution, and one should
have a large computation time to see the convergence. The top
figures in Figure \ref{fig:X-i} show the numerical simulation.
 The simulation indicates the generation of the phase shift
in the lower part of the solution. This may be explained as follows:  Since the $[1,4]$-soliton propagates faster than the $[2,3]$-soliton, the intersection point
moves in the upper direction, i.e. the positive $y$-direction.
 Then the effect of the nonlinear interaction may be observed in the lower half (i.e.
 behind the interactions) of the solution, and it generates the phase shifts.
 One should also note that the large bending of the $[1,4]$-soliton in $y<0$. This is due to the interaction
 with the dispersive wake generated around the origin at $t=0$, and notice that this bending part disperses
 away. The middle figures in Figure \ref{fig:X-i} show the corresponding exact solution of P-type
 whose $A$-matrix is obtained by minimizing the error function $E(t)$ of \eqref{error} at $t=12$,
 \[
A=\begin{pmatrix}
1 &    0   &  0  & -0.038  \\
0 &   1  & 1.629  &     0
\end{pmatrix}
\]
In the minimization, we adjust the location of the line-solitons at
the wavefront, and using (\ref{PA}), we obtain
\[
x_{[2,3]}^+=-0.125,\qquad x_{[1,4]}^+=-0.54.
\]
Notice that the larger shifts in the $[1,4]$-soliton, which can be also observed in the simulation.
This shift is due to the repulsive force appearing at the intersection point (recall that the exact solution has
a smaller amplitude at this point). The phase shifts of the P-type solution is given by (\ref{Pshift}),
\[
\Delta x_{[1,4]}=-1.10,\qquad \Delta x_{[2,3]}=-2.75.
\]
Those large phase shifts are also observed in the simulation, and the bending regions
are getting separated from the intersection point with the correct phase shifts.
The bottom graph in Figure \ref{fig:X-i} shows the decay of the error function $E(t)$
with $D_r^t$ for $r=10$.
We also note the saturation $E(t)$ after $t=11$, which may be due to the large deformation
of the solution, and we may need to perform a larger domain computation.

\subsubsection{The case (j)} This case is similar to the previous one with
 $[1,4]$-  and $[2,3]$-soliton in the initial wave.   However in this case, the larger
 soliton of $[1,4]$-type has a negative angle (propagating in the negative $y$-direction),
 and we expect that the phase shift now appear in the upper half of the solution.
 We take the amplitudes and angles
of those solitons to be
\[\left\{
\begin{array}{llll}
A_{[1,4]}=A_0=\frac{169}{32},  &  \tan\Psi_{[1,4]}=-\frac{1}{4}=-\tan\Psi_0\\[1.0ex]
A_{[2,3]}=2 & \tan\Psi_{[2,3]}=\frac{1}{4}=\tan\Psi_0
\end{array}
\right.
\]
The $k$-parameters are then given by $(k_1,k_2,k_3,k_4)=(-\frac{7}{4},
-\frac{7}{8},\frac{9}{8},\frac{3}{2})$.
%%%%%%%%%%%%%%%%%%%%%%%%%%%%%%%%%%%%%
\begin{figure}[t!]
\begin{centering}
\includegraphics[scale=0.63]{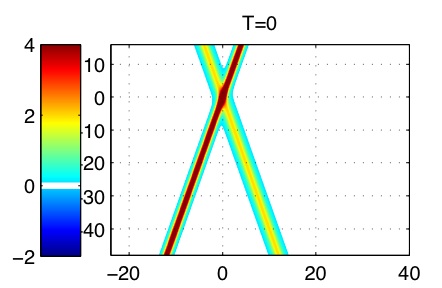}\includegraphics[scale=0.6]{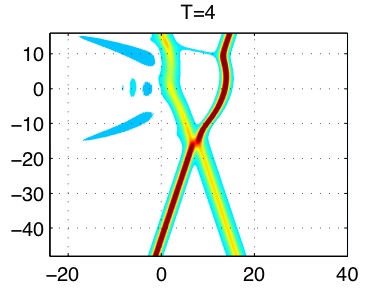}\includegraphics[scale=0.6]{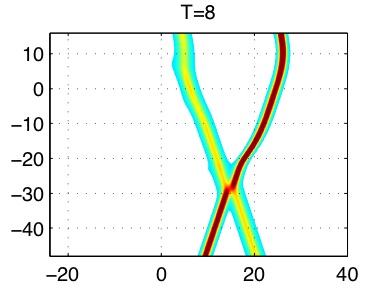}
\par\end{centering}
\begin{centering}
\includegraphics[scale=0.63]{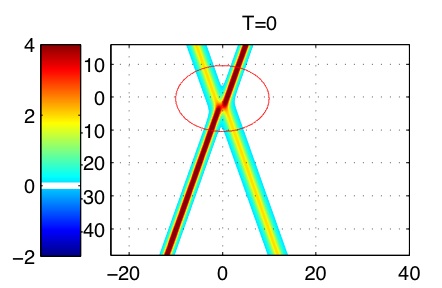}\includegraphics[scale=0.6]{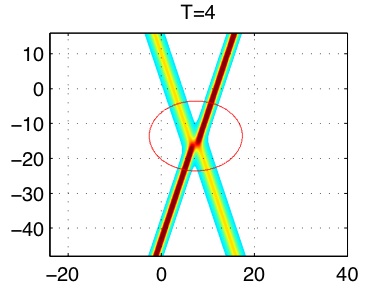}\includegraphics[scale=0.6]{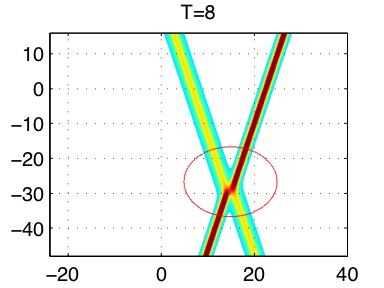}
\par\end{centering}
\begin{centering}
\includegraphics[scale=0.7]{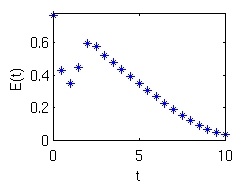}
\par\end{centering}
\caption{\label{fig:X-j} Numerical simulation of the case (j) in
Figure \ref{fig:X}. The initial wave is the sum of $[1,4]$- and
$[2,3]$-solitons.  The top figures show the numerical simulation,
and the middle figures show the corresponding exact solution of
$(4321)$-type, i.e. P-type. The bottom graph illustrates the error
function $E(t)$ of \eqref{error} which is minimized at $t=8$. The
graph shows a further reduction of the error, that is, the
convergence of the solution to the corresponding exact solution even
with a drastic change of the error at the initial stage.}
\end{figure}
%%%%%%%%%%%%%%%%%%%%%%%%%%%%%%%%%%%%%%%%%%%
The top figures in Figure \ref{fig:X-j} show the numerical
simulation. One can observe a dispersive wake is generated by a
large perturbation in the initial wave near the origin, and then the
wake pushes the $[1,4]$-soliton forward. Also the large bending of
the $[2,3]$-soliton near the origin is due to the interaction with
the wake. Since the $[1,4]$-soliton propagates much faster than the
other one, the intersection point moves in the negative
$y$-direction. We can see a clear separation between the dispersive
waves causing the deformation of solitons and the intersection
points of P-type. The middle figures in Figure \ref{fig:X-j}
illustrate the corresponding P-type solution whose $A$-matrix is
obtained by minimizing the error function $E(t)$ at $t=8$. We take
the radius $r=10$ for a good separation of the dispersive
perturbation from the steady P-type soliton part. With this
minimization, we get
\[
A=\begin{pmatrix}
1 &    0   &  0  &   -0.299 \\
0 &   1  &   8.549 &     0
\end{pmatrix}
\]
In the minimization, we adjust the locations of the line-solitons at
the wavefront, and using (\ref{PA}), we obtain
\[
x_{[1,4]}^+=0.627,\qquad x_{[2,3]}^+=0.923.
\]
The positive shifts are due to the repulsive interaction at the intersection point as observed
in the simulation.
The phase shifts of those incidence solitons are given by (\ref{Pshift}),
\[
\Delta x_{[1,4]}=-0.934,\qquad \Delta x_{[2,3]}=-1.518.
\]
Notice that the $[1,4]$-soliton behind the $[2,3]$-soliton shifts $x^-_{[1,4]}=x^+_{[1,4]}+\Delta x_{[1,4]}=-0.307$, and it gets a larger shift
in the upper part. The large bending in the simulation is generated by a strong perturbation generated
at the intersection point at $t=0$. However, the bending part disperses and the intersection point
gets away from this part.

The bottom graph in Figure \ref{fig:X-j} shows the decay of the error.  We emphasize that the error function
gets a rapid change in the early times. This implies that a drastic change in the profile of the solution
due to a large perturbation to the exact solution. However after $t=3$ the solution has a monotone
convergence to the exact solution, implying that the dispersive waves are getting away from the
domain $D_r^t$ as we expected.

%
%\begin{figure}[t!]
%\begin{centering}
%\includegraphics[scale=0.5]{Sol33_0_0/Sol33_0_0_u_0}\includegraphics[scale=0.5]{Sol33_0_0/Sol33_0_0_u_3}\includegraphics[scale=0.5]{Sol33_0_0/Sol33_0_0_u_6}
%\par\end{centering}

%\begin{centering}
%\includegraphics[scale=0.5]{Sol33_0_0/exact_33soliton_0}\includegraphics[scale=0.5]{Sol33_0_0/exact_33soliton_3}\includegraphics[scale=0.5]{Sol33_0_0/exact_33soliton_6}
%\par\end{centering}

%\caption{\label{fig: 415362}(a)$[k_{1},k_{2},k_{3},k_{4},k_{5},k_{6}]=[-2,-3/2,-1,0,1/2,2]$.
%$ $$[A_{[1,2]},A_{[3,4]},A_{[2,6]},A_{[5,6]},A_{[3,5]},A_{[1,4]}]=[0.25,0.5,6.125,1.125,1.125,2]$,$[\tan(\Psi_{[1,2]}),\tan(\Psi_{[3,4]}),\tan(\Psi_{[2,6]}),\tan(\Psi_{[5,6]}),\tan(\Psi_{[3,5]}),\tan(\Psi_{[1,4]})]=[-\frac{7}{2},-1,\frac{1}{2},\frac{3}{2},1,-2]$}

%\end{figure}

\vskip 0.5cm

\leftline{\bf Acknowledgement}
\bigskip

\noindent We would like to thank Professor Thiab Taha for giving us
the opportunity to present our result at the Sixth IMACS
International Conference on March 23-26, 2009. We would also like to
thank Professors Masayuki Oikawa, Hidekazu Tsuji and Sarvarish
Chakravarty for various useful discussions related to the present
work.  C.-Y. K is partially supported by NSF grant DMS-0811003 and
the Alfred P. Sloan fellowship. Y.K. is partially supported by NSF
grant DMS-0806219.

%%%%%%%%%%%%%%%%%%%%%%%%%%%%%%%%%%%%%%%%%%

\end{document}